\documentclass[11pt]{article}
\usepackage{graphicx} 
\usepackage[utf8]{inputenc}
\usepackage{titlesec}
\titleformat{\section}
{\normalfont\scshape}
{\thesection.}{0.5em}{\centering}
\titleformat{\subsection}
{\small\scshape}
{\thesubsection.}{0.5em}{\centering}
\titleformat{\subsubsection}
{\small\scshape}
{\thesubsubsection.}{0.5em}{\centering}
\usepackage{tikz-cd}
\usepackage{tikz}
\usepackage[left=3.5cm,right=3.5cm,top=3.5cm,bottom=3.5cm]{geometry}
\usepackage{comment}
\usepackage{amsfonts}
\usepackage{indentfirst}
\renewcommand{\abstract}{\small{\section*{\abstractname}}}
\usepackage{hyperref}
\hypersetup{
    colorlinks,
    citecolor=teal,
    filecolor=teal,
    linkcolor=teal,
    urlcolor=teal
}
\usepackage{amsmath}
\usepackage{amssymb}
\usepackage{amsthm}
\newtheorem*{theorem*}{Theorem}
\usepackage{mathrsfs}
\newtheorem{theorem}{Theorem}
\newtheorem{mainthm}{Theorem}

\newtheorem{lemma}{Lemma}
\newtheorem{proposition}{Proposition}
\newtheorem{corollary}{Corollary}
\newtheorem{definition}{Definition}
\theoremstyle{remark}
\newtheorem{remark}{Remark}
\newtheorem{example}{Example}

\usepackage{stmaryrd}
\usepackage{enumitem}

\usepackage[
backend=biber,
style=alphabetic,
sorting=ynt
]{biblatex}
\title{\Large{\textbf{Optimal Risk Sharing Without Preference Convexity: An Aggregate Convexity Approach}}}
\author{\large{Vasily Melnikov}}
\date{\normalsize{\textsc{August 2026}}}

\begin{document}

\maketitle
\begin{abstract}
     We consider the optimal risk sharing problem with a continuum of agents, modeled via a non-atomic measure space. Individual preferences are not assumed to be convex. We show the multiplicity of agents induces the value function to be convex, allowing for the application of convex duality techniques to risk sharing without preference convexity. A computationally tractable formula for the conjugate of the value function is derived, yielding an explicit dual representation of the value function. Applications of our results include a version of the two fundamental theorems of welfare economics for a large class of non-convex preferences, and non-existence results for Pareto optima when preferences are distortion risk measures whose corresponding distortion function fails to majorize the identity.
\end{abstract}
\section{Introduction}\label{sec:intro}
An individual will aim to minimize risk, and consequently may share risk with others as a means to that end. Given $N$ agents and their risk preferences, represented by a sequence of risk measures $(\varrho_{i})_{i=1}^{N}$, the optimal risk sharing problem for a risk $\mathcal{X}$ to distribute is
\begin{equation}\label{eq:min-problem}
    \sum_{i=1}^{N}\varrho_{i}(X_{i})\longrightarrow\textrm{min!}
\end{equation}
subject to the constraint $\sum_{i=1}^{N}X_{i}=\mathcal{X}$, where $(X_{i})_{i=1}^{N}$ and $\mathcal{X}$ are bounded random variables on a probability space $(\Omega,\mathscr{F},\mathbb{P})$. For a social planner, (\ref{eq:min-problem}) functions either as a mandate or as the reaction of agents to regulation; see Weber \cite{solvencyiibad}, Filipović and Kupper \cite{solvencyiimf}, and Liebrich and Svinland \cite{caprequirfs}. Understood otherwise, (\ref{eq:min-problem}) characterizes the Pareto efficient allocations, reformulating a fundamental question in welfare economics into a tractable optimization problem; see Ghossoub and Zhu \cite{ghossoubzhu2025}.
\par
A significant literature studies risk sharing when preferences are convex, which allows the application of convex analytic tools; see Heath and Ku \cite{heath-ku}, Acciaio \cite{dualnonmonotone}, and Jouini, Schachermayer, and Touzi \cite{law-inv-util-conv-dual}. Without convexity, convex analysis has a more limited applicability, so a replacement methodology is needed. To that end, Mao and Wang \cite{consistency} and Liebrich \cite{felixfs} apply comonotonicity to the consistent risk sharing problem, while Embrechts, Liu, and Wang \cite{quantile-share} develop \textit{ad hoc} techniques for quantile-based risk sharing. We depart from existing methodologies for non-convex risk sharing, and instead consider the effect that a large number of agents has on (\ref{eq:min-problem}).
\par
Our approach is rooted in the observation, originally by Edgeworth \cite{edge-psychic}, that a large number of agents regularizes an economy. For example, sets formed by aggregating individual preferences become convex or close to convex as the number of agents increases, irrelevant of the shape of individual preferences, an effect first identified explicitly by Farrell \cite[p. 381]{convexity-assumption}.
\par
The consequent ability to apply separation theorems, among other tools relying on convexity, allows one to prove strong regularity results for large economies which may fail individual preference convexity. We list some examples. First, the core often coincides with the set of competitive equilibria; see Aumann \cite{aumanncontinuum}, Vind \cite{vind-edge}, and Tourky and Yannelis \cite{hidden-assumption}. Second, competitive equilibria often exist; see Aumann \cite{aumann66}, Starr \cite{starr}, Lee \cite{lee-comp}, and Jang and Lee \cite{jang-lee}. Third, the welfare theorems often hold; see Farrell \cite{convexity-assumption}, Hildenbrand \cite{hild-welfare}, and Anderson \cite{welfare-theorem-second-anderson}. When agents are few or finite, the aforementioned results either do not hold, or otherwise require convexity of individual preferences.
\par
We find similar regularizing effects when the risk sharing problem (\ref{eq:min-problem}) involves a continuum of agents, which we model using the measure space risk sharing framework of \cite{melnikov-general-theory}. Agents are represented by the points of a complete non-atomic measure space $(A,\mathscr{A},\mu)$, and each agent $a\in A$ has an associated risk measure $\varrho_{a}$ on $L^{\infty}(\mathbb{P})$; no convexity assumptions on $\varrho_{a}$ are made, although the preference mapping $a\longmapsto\varrho_{a}$ is required to be measurable in an appropriate sense (see Definition \ref{def:a-meas-pref}). For a risk $\mathcal{X}$ to distribute, the problem (\ref{eq:min-problem}) becomes
\begin{equation}\label{eq:min-problem-measure}
    \int_{A}\varrho_{a}(X_{a})\mu(da)\longrightarrow\textrm{min!}
\end{equation}
subject to the Gelfand integral $\int_{A}X_{a}\mu(da)$ existing and equaling $\mathcal{X}$. Denoting by $\mathcal{X}\longmapsto\left(\operatorname{\Box}_{a\in A}\varrho_{a}\mu(da)\right)(\mathcal{X})$ the value function of (\ref{eq:min-problem-measure}), we have the following convexification result.
\begin{mainthm}\label{thm:intro-convexify}
    Under general conditions, $\operatorname{\Box}_{a\in A}\varrho_{a}\mu(da)$ is a convex risk measure.
\end{mainthm}
Theorem \ref{thm:intro-convexify} is established under four different sets of assumptions (for each set, we list here only the most limiting assumption of the set). The first set requires the $\sigma$-algebra $\mathscr{F}$ to be finite (see Theorem \ref{thm:finite-conv}). The second set requires the individual preferences $\varrho_{a}$ to be positively homogeneous (see Theorem \ref{thm:coherent}). The third set requires the individual preferences $\varrho_{a}$ to satisfy consistency in the sense of Mao and Wang \cite{consistency}, a mild risk aversion condition (see Theorem \ref{thm:general-conv}). The fourth set requires $\mu$ to be saturated, a strengthened form of non-atomicity (see Proposition \ref{prop:saturated}). In each of these listed cases, convexity implies the value function admits a dual representation, for which we provide an explicit formula (see Theorem \ref{thm:conv-conj-value}).
\par
The regularity properties Theorem \ref{thm:intro-convexify} establishes for the value function are often sufficient to conclude that the subdifferential is non-empty at all points. Interpreting the value function of (\ref{eq:min-problem-measure}) as a social welfare function, the subdifferential corresponds to pricing vectors for competitive equilibria. As an upshot of this observation, we have the following version of the fundamental theorems of welfare economics.
\begin{mainthm}\label{thm:intro-welfare-theorems}
    Suppose each $\varrho_{a}$ has the Lebesgue property, and $\operatorname{\Box}_{a\in A}\varrho_{a}\mu(da)$ is a convex risk measure. Then:
    \begin{enumerate}
        \item If $(X_{a})_{a\in A}$ is a competitive equilibrium under some pricing measure, then $(X_{a})_{a\in A}$ is Pareto efficient.
        \item If $(X_{a})_{a\in A}$ is Pareto efficient, there exists a real-valued $(C_{a})_{a\in A}$ with $\int_{A}C_{a}\mu(da)=0$ such that $(X_{a}+C_{a})_{a\in A}$ is a competitive equilibrium under some pricing measure.
    \end{enumerate}
\end{mainthm}
Instead of requiring convexity of individual preferences, as in the classical setting, Theorem \ref{thm:intro-welfare-theorems} merely requires that the value function is convex—guaranteed to hold in many situations by Theorem \ref{thm:intro-convexify}.
\par
Theorem \ref{thm:intro-welfare-theorems} is reproduced in the main text as Theorem \ref{thm:welfare-theorems}; applications of the result to particular classes of preferences, such as to consistent or positively homogeneous preferences, can be found in Corollary \ref{corr:second-welfare-thm-consistent} and Corollary \ref{corr:second-welfare-positive-homogen}.
\par
Improperness of the value function implies the non-existence of Pareto optima when there are finitely many agents. If improperness is strengthened to a hereditary condition (see Definition \ref{def:hereditary-not-posed}), a similar principle holds for a continuum of agents (see Theorem \ref{thm:no-pareto-ill-posed}). As an application, we obtain non-existence results for Pareto optima whenever each $\varrho_{a}$ is a distortion risk measure whose associated distortion function fails to majorize the identity (see Corollary \ref{corr:cong-degen-pareto}). This class of distortion risk measures is vast; for example, it includes the empirically realistic risk preferences introduced by Tversky and Kahneman \cite{tk}.
\par
The paper is structured as follows. In \S\ref{sec:prelim}, we establish our basic framework. In \S\ref{sec:finite-prob-space-conv}, \S\ref{sec:coherent}, and \S\ref{sec:conv-non-atom}, we show the value function of the risk sharing problem is convex under various assumptions on the structure of preferences, and a dual representation of the value function is furnished in \S\ref{sec:conv-conj-value}. The properness of the value function, a question closely related to the aforementioned convexification results, is studied in \S\ref{sec:improper}. In \S\ref{sec:pareto-equi}, economic applications of the convexification and improperness results are obtained, including non-existence results for Pareto optima and versions of the welfare theorems.
\section{The Basic Framework}\label{sec:prelim}
In this section, we recall the optimal risk sharing framework proposed by \cite{melnikov-general-theory}, which we use in the sequel; the setting is different, since we allow non-convex risk measures, but the definitions are essentially unchanged from the convex case.
\par
Agents are represented by a complete measure space $(A,\mathscr{A},\mu)$, where $0<\mu(A)<\infty$. Unless otherwise stated, it is assumed that $(A,\mathscr{A},\mu)$ is non-atomic. We extend the definition of the Lebesgue integral beyond $\mu$-integrable functions by setting
\begin{equation}\label{eq:extended-lebesgue-int}
    \int_{A}\xi(a)\mu(da)=\inf_{n\in\mathbb{N}}\int_{A}(-n)\vee\xi(a)\mu(da)\in\mathbb{R}\cup\{\pm\infty\}
\end{equation}
for any $\mathscr{A}$-measurable function $\xi:A\longrightarrow\mathbb{R}\cup\{\pm\infty\}$.
\par
Uncertainty is modeled by a separable probability space $(\Omega,\mathscr{F},\mathbb{P})$; $\Omega$ represents the possible states of the world, while the $\sigma$-algebra $\mathscr{F}$ consists of all events discernible from the \textit{ex post} available information about $\Omega$.
\par
The spaces $L^{1}(\mathbb{P})$ and $L^{\infty}(\mathbb{P})$ carry their usual meaning as spaces of contingent payoffs, although we adopt the convention that $\mathcal{X}\geq0$ represents a loss of magnitude $\mathcal{X}$. With $\mathscr{M}_{\mathbb{P}}$ we denote the set of absolutely continuous probability measures $\mathbb{Q}\ll\mathbb{P}$ on $\mathscr{F}$; for notational convenience, $\mathscr{M}_{\mathbb{P}}$ will sometimes be viewed as a subset of $L^{1}(\mathbb{P})$ via the Radon-Nikodým derivative.
\subsection{Allocations}
A function $a\longmapsto X_{a}$ from $A$ to $L^{\infty}(\mathbb{P})$, assigning to each agent $a\in A$ some contingent payoff $X_{a}\in L^{\infty}(\mathbb{P})$, is called an allocation. Applying an integration theory to such functions requires making measurability assumptions.
\begin{definition}\label{def:meas-weak-star}
    An allocation $(X_{a})_{a\in A}$ is said to be $\mathscr{A}$-measurable if, for each $\mathcal{Y}\in L^{1}(\mathbb{P})$, the function $a\longmapsto\mathbb{E}^{\mathbb{P}}(X_{a}\mathcal{Y})$ is $\mathscr{A}$-measurable.
\end{definition}
Equipped with the above notion, we may define an integration theory for allocations.
\begin{definition}\label{def:gelf-int}
    An $\mathscr{A}$-measurable allocation $(X_{a})_{a\in A}$ is said to be Gelfand integrable if, for each $\mathcal{Y}\in L^{1}(\mathbb{P})$, the $\mathscr{A}$-measurable function $a\longmapsto\mathbb{E}^{\mathbb{P}}(X_{a}\mathcal{Y})$ is $\mu$-integrable.
\end{definition}
If $(X_{a})_{a\in A}$ is Gelfand integrable, for each $B\in\mathscr{A}$, there exists a unique element $\mathcal{Z}_{B}\in L^{\infty}(\mathbb{P})$ such that
\begin{equation*}
        \mathbb{E}^{\mathbb{P}}\left(\mathcal{Z}_{B}\mathcal{Y}\right)=\int_{B}\mathbb{E}^{\mathbb{P}}\left(X_{a}\mathcal{Y}\right)\mu(da),
\end{equation*}
for each $\mathcal{Y}\in L^{1}(\mathbb{P})$ (see \cite[p. 430]{aliprantis-inf-dim}). We call $\mathcal{Z}_{B}$ the Gelfand integral of $(X_{a})_{a\in A}$ over $B$, and write $\mathcal{Z}_{B}=\int_{B}X_{a}\mu(da)$.
\par
As a generalization of the feasibility constraint from (\ref{eq:min-problem}), given a risk $\mathcal{X}\in L^{\infty}(\mathbb{P})$, we constrain any allocation to satisfy the following.
\begin{definition}
    An $\mathscr{A}$-measurable allocation $(X_{a})_{a\in A}$ is said to be $\mathcal{X}$-feasible if $(X_{a})_{a\in A}$ is Gelfand integrable, and $\mathcal{X}=\int_{A}X_{a}\mu(da)$. The set of such allocations is denoted $\mathbb{A}(\mathcal{X})$:
    \begin{equation*}
        \mathbb{A}(\mathcal{X})=\left\{(X_{a})_{a\in A}:(X_{a})_{a\in A}\textrm{ is Gelfand integrable and }\int_{A}X_{a}\mu(da)=\mathcal{X}\right\}.
    \end{equation*}
\end{definition}
\subsection{Risk Measures}
To model the individual preferences of agents, we use risk measures; the relation of risk measures to risk preferences is explained in \S\ref{subsec:pref}. A risk measure is a functional $\varrho:L^{\infty}(\mathbb{P})\longrightarrow\mathbb{R}$ satisfying properties (\ref{it:monotonicity}) and (\ref{it:cash-add}); $\varrho$ is said to be a convex (respectively, coherent) risk measure if $\varrho$ satisfies (\ref{it:monotonicity}) to (\ref{it:convexity}) (respectively, (\ref{it:monotonicity}) to (\ref{it:pos-hom})).
\begin{enumerate}[label=\upshape\Roman*.,ref=\upshape\Roman*]
    \item\label{it:monotonicity} Monotonicity: for each $\mathcal{X},\mathcal{Y}\in L^{\infty}(\mathbb{P})$, if $\mathcal{X}\geq\mathcal{Y}$, then $\varrho(\mathcal{X})\geq\varrho(\mathcal{Y})$.
    \item\label{it:cash-add} Cash additivity: for each $\mathcal{X}\in L^{\infty}(\mathbb{P})$, if $c\in\mathbb{R}$, then $\varrho(\mathcal{X}+c)=\varrho(\mathcal{X})+c$.
    \item\label{it:convexity} Convexity: for each $\mathcal{X},\mathcal{Y}\in L^{\infty}(\mathbb{P})$ and $\lambda\in[0,1]$, $\varrho(\lambda\mathcal{X}+(1-\lambda)\mathcal{Y})\leq\lambda\varrho(\mathcal{X})+(1-\lambda)\varrho(\mathcal{Y})$.
    \item\label{it:pos-hom} Positive Homogeneity: for each $\mathcal{X}\in L^{\infty}(\mathbb{P})$ and $\lambda\in(0,\infty)$, $\varrho(\lambda\mathcal{X})=\lambda\varrho(\mathcal{X})$.
    \item\label{it:fatou} Fatou property: if $(\mathcal{X}^{n})_{n=1}^{\infty}\subseteq L^{\infty}(\mathbb{P})$ is an $L^{\infty}(\mathbb{P})$-bounded sequence converging in probability to $\mathcal{X}\in L^{\infty}(\mathbb{P})$, then
    \begin{equation*}
        \varrho(\mathcal{X})\leq\liminf_{n\to\infty}\varrho(\mathcal{X}^{n}).
    \end{equation*}
    \item\label{it:lebesgue} Lebesgue property: if $(\mathcal{X}^{n})_{n=1}^{\infty}\subseteq L^{\infty}(\mathbb{P})$ is an $L^{\infty}(\mathbb{P})$-bounded sequence converging in probability to $\mathcal{X}\in L^{\infty}(\mathbb{P})$, then $\lim_{n\to\infty}\varrho(\mathcal{X}^{n})$ exists and equals $\varrho(\mathcal{X})$.
    \item\label{it:cont-from-above} Continuity from above: if $(\mathcal{X}^{n})_{n=1}^{\infty}\subseteq L^{\infty}(\mathbb{P})$ is a decreasing $L^{\infty}(\mathbb{P})$-bounded sequence with $\mathbb{P}$-a.s. limit $\mathcal{X}=\inf_{n}\mathcal{X}^{n}$, then $\lim_{n\to\infty}\varrho(\mathcal{X}^{n})$ exists and equals $\varrho(\mathcal{X})$.
    \item\label{it:cont-from-below} Continuity from below: if $(\mathcal{X}^{n})_{n=1}^{\infty}\subseteq L^{\infty}(\mathbb{P})$ is an increasing $L^{\infty}(\mathbb{P})$-bounded sequence with $\mathbb{P}$-a.s. limit $\mathcal{X}=\sup_{n}\mathcal{X}^{n}$, then $\lim_{n\to\infty}\varrho(\mathcal{X}^{n})$ exists and equals $\varrho(\mathcal{X})$.
\end{enumerate}
As defined above, a risk measure is a monotone cash additive functional. In the literature, monotone cash additive functionals are sometimes instead called monetary risk measures.
\par
The Fatou property and the Lebesgue property can be reformulated in terms of continuity from above or below. More precisely, a risk measure $\varrho$ has the Fatou property iff $\varrho$ is continuous from below, while a risk measure $\varrho$ has the Lebesgue property iff $\varrho$ is continuous from below and above.
\par
As a consequence of cash additivity, any risk measure $\varrho$ can be identified with its acceptance set $\mathfrak{A}(\varrho)=\{\mathcal{X}\in L^{\infty}(\mathbb{P}):\varrho(\mathcal{X})\leq0\}$ via the formula
\begin{equation*}
    \varrho(\mathcal{X})=\inf\left\{m\in\mathbb{R}:\mathcal{X}-m\in\mathfrak{A}(\varrho)\right\}
\end{equation*}
for each $\mathcal{X}\in L^{\infty}(\mathbb{P})$.
\par
Let $\varrho:L^{\infty}(\mathbb{P})\longrightarrow\mathbb{R}\cup\{\pm\infty\}$ be any functional. The convex conjugate $\varrho^{\ast}:L^{1}(\mathbb{P})\longrightarrow\mathbb{R}\cup\{\pm\infty\}$ of $\varrho$ is defined by
\begin{equation*}
    \forall\mathcal{Y}\in L^{1}(\mathbb{P}):\varrho^{\ast}(\mathcal{Y})=\sup_{\mathcal{X}\in L^{\infty}(\mathbb{P})}\left(\mathbb{E}^{\mathbb{P}}(\mathcal{X}\mathcal{Y})-\varrho(\mathcal{X})\right).
\end{equation*}
If $\varrho$ is a risk measure, then $\varrho^{\ast}(\mathcal{Y})<\infty$ implies that $\mathcal{Y}$ is the $\mathbb{P}$-density $\frac{d\mathbb{Q}}{d\mathbb{P}}$ of some $\mathbb{Q}\in\mathscr{M}_{\mathbb{P}}$. By an abuse of notation, $\varrho^{\ast}$ can take probability measures as inputs: if $\mathbb{Q}\in\mathscr{M}_{\mathbb{P}}$, we write $\varrho^{\ast}(\mathbb{Q})$ for $\varrho^{\ast}(\frac{d\mathbb{Q}}{d\mathbb{P}})$.
\par
If $\varrho$ is a convex risk measure with the Fatou property, we have the dual representation
\begin{equation}\label{eq:dual-rep}
    \varrho(\mathcal{X})=\sup_{\mathcal{Y}\in L^{1}(\mathbb{P})}\left(\mathbb{E}^{\mathbb{P}}(\mathcal{X}\mathcal{Y})-\varrho^{\ast}(\mathcal{Y})\right)=\sup_{\mathbb{Q}\in\mathscr{M}_{\mathbb{P}}}\left(\mathbb{E}^{\mathbb{Q}}(\mathcal{X})-\varrho^{\ast}(\mathbb{Q})\right)
\end{equation}
for each $\mathcal{X}\in L^{\infty}(\mathbb{P})$. In lieu of a representation of the form (\ref{eq:dual-rep}) for a general functional $\varrho$, we define the biconjugate $\varrho^{\ast\ast}$ for functionals $\varrho:L^{\infty}(\mathbb{P})\longrightarrow\mathbb{R}\cup\{\pm\infty\}$ by
\begin{equation*}
    \varrho^{\ast\ast}(\mathcal{X})=\sup_{\mathcal{Y}\in L^{1}(\mathbb{P})}\left(\mathbb{E}^{\mathbb{P}}(\mathcal{X}\mathcal{Y})-\varrho^{\ast}(\mathcal{Y})\right)
\end{equation*}
for each $\mathcal{X}\in L^{\infty}(\mathbb{P})$. If $\varrho$ is a risk measure, the above supremum can be taken over $\mathscr{M}_{\mathbb{P}}$ rather than all of $L^{1}(\mathbb{P})$.
\subsection{Law Invariance and Consistency}
Let $\mathbb{Q}\ll\mathbb{P}$ be a probability measure. A risk measure $\varrho$ is said to be $\mathbb{Q}$-law invariant if $\mathbb{Q}\circ\mathcal{X}^{-1}=\mathbb{Q}\circ\mathcal{Y}^{-1}$ implies $\varrho(\mathcal{X})=\varrho(\mathcal{Y})$. For non-convex risk measures, law invariance is often not a strong enough property. As an alternative, we use the notion of consistency, introduced by Mao and Wang \cite{consistency}.
\begin{definition}
    Let $\mathbb{Q}\in\mathscr{M}_{\mathbb{P}}$. For random variables $\mathcal{X},\mathcal{Y}\in L^{\infty}(\mathbb{P})$, we say that $\mathcal{X}$ is second order $\mathbb{Q}$-stochastic dominated by $\mathcal{Y}$, denoted $\mathcal{X}\lesssim_{c,\mathbb{Q}}\mathcal{Y}$, if $\mathbb{E}^{\mathbb{Q}}(\varphi(\mathcal{X}))\leq\mathbb{E}^{\mathbb{Q}}(\varphi(\mathcal{Y}))$ for all increasing convex functions $\varphi:\mathbb{R}\longrightarrow\mathbb{R}$.
\end{definition}
\begin{definition}
    Let $\mathbb{Q}\in\mathscr{M}_{\mathbb{P}}$. A risk measure $\varrho$ is said to be $\mathbb{Q}$-consistent if $\varrho$ preserves second order $\mathbb{Q}$-stochastic dominance. More precisely, $\varrho$ is $\mathbb{Q}$-consistent if
    \begin{equation*}
        \mathcal{X}\lesssim_{c,\mathbb{Q}}\mathcal{Y}\implies\varrho(\mathcal{X})\leq\varrho(\mathcal{Y}).
    \end{equation*}
\end{definition}
Mao and Wang \cite{consistency} establish several equivalent characterizations and properties of consistent risk measures. Of particular importance to us is the relation of consistent risk measures to dilatation monotone risk measures, defined as follows.
\begin{definition}
    Let $\mathbb{Q}\sim\mathbb{P}$. A risk measure $\varrho$ is said to be $\mathbb{Q}$-dilatation monotone if
    \begin{equation*}
        \varrho(\mathbb{E}^{\mathbb{Q}}(\mathcal{X}|\mathscr{G}))\leq\varrho(\mathcal{X})
    \end{equation*}
    for every $\mathcal{X}\in L^{\infty}(\mathbb{P})$ and sub-$\sigma$-algebra $\mathscr{G}\subseteq\mathscr{F}$.
\end{definition}
\begin{proposition}\label{prop:dm-iff-cons}
    Let $\mathbb{Q}\sim\mathbb{P}$ be non-atomic. A risk measure $\varrho$ is $\mathbb{Q}$-dilatation monotone iff $\varrho$ is $\mathbb{Q}$-consistent.
\end{proposition}
\begin{proof}
    This is the content of \cite[Theorem 2.1]{consistency}.
\end{proof}
A classical result of Jouini, Schachermayer, and Touzi \cite{schtouz} is that a law invariant convex risk measure automatically satisfies the Fatou property. Similarly, without convexity, one has the following result for consistent risk measures, which is motivated by the fact that consistency implies law invariance (see \cite[Theorem 2.1]{consistency}).
\begin{proposition}\label{prop:cons-fatou}
    Let $\mathbb{Q}\sim\mathbb{P}$ be non-atomic. If $\varrho$ is $\mathbb{Q}$-consistent, $\varrho$ satisfies the Fatou property.
\end{proposition}
\begin{proof}
    This is the content of \cite[Theorem 3.5]{consistency}.
\end{proof}
Mao and Wang \cite{consistency} show that $\varrho$ is $\mathbb{Q}$-consistent iff
\begin{equation}\label{eq:consistent-rep}
    \forall\mathcal{X}\in L^{\infty}(\mathbb{P}):\varrho(\mathcal{X})=\inf_{\tau\in\mathcal{T}}\varrho_{\tau}(\mathcal{X})
\end{equation}
for some collection $\{\varrho_{\tau}:\tau\in\mathcal{T}\}$ of $\mathbb{Q}$-law invariant convex risk measures. A direct consequence of (\ref{eq:consistent-rep}) is a formula for the conjugate of $\varrho$:
\begin{equation*}
    \forall\mathbb{S}\in\mathscr{M}_{\mathbb{P}}:\varrho^{\ast}(\mathbb{S})=\sup_{\tau\in\mathcal{T}}\varrho^{\ast}_{\tau}(\mathbb{S}).
\end{equation*}
We often assume the Lebesgue property in the sequel, so it behooves us to note a simple way to check the Lebesgue property of a consistent risk measure in terms of any representation of the form (\ref{eq:consistent-rep}).
\begin{proposition}\label{prop:checking-lebesgue-consistent}
    Let $\mathbb{Q}\sim\mathbb{P}$. Suppose $\varrho$ is a $\mathbb{Q}$-consistent risk measure satisfying (\ref{eq:consistent-rep}) for some collection $\{\varrho_{\tau}:\tau\in\mathcal{T}\}$ of $\mathbb{Q}$-law invariant convex risk measures. If $\varrho_{\tau}$ has the Lebesgue property for each $\tau\in\mathcal{T}$, then $\varrho$ has the Lebesgue property.
\end{proposition}
\begin{proof}
    As $\varrho$ is $\mathbb{Q}$-consistent, Proposition \ref{prop:cons-fatou} implies $\varrho$ has the Fatou property, so $\varrho$ is continuous from below. To show $\varrho$ has the Lebesgue property, it thus suffices to show $\varrho$ is continuous from above.
    \par
    Let $(\mathcal{X}^{n})_{n=1}^{\infty}\subseteq L^{\infty}(\mathbb{P})$ be $L^{\infty}(\mathbb{P})$-bounded and decreasing to $\mathcal{X}\in L^{\infty}(\mathbb{P})$. Then
    \begin{equation*}
        \inf_{n}\varrho(\mathcal{X}^{n})=\inf_{n}\inf_{\tau\in\mathcal{T}}\varrho_{\tau}(\mathcal{X}^{n})=\inf_{\tau\in\mathcal{T}}\inf_{n}\varrho_{\tau}(\mathcal{X}^{n})=\inf_{\tau\in\mathcal{T}}\varrho_{\tau}(\mathcal{X})=\varrho(\mathcal{X}).
    \end{equation*}
    since each $\varrho_{\tau}$ has the Lebesgue property. So, $\varrho$ is continuous from above, as desired.
\end{proof}
\subsection{Risk Preferences}\label{subsec:pref}
Each agent has risk preferences, which are modeled by a risk measure. For each agent $a\in A$, we therefore have a risk measure $\varrho_{a}$, representing the risk preferences of the agent: $\mathcal{X}$ is weakly preferred to $\mathcal{Y}$ by agent $a$ if $\varrho_{a}(\mathcal{X})\leq\varrho_{a}(\mathcal{Y})$. Collecting all of the preferences yields a collection $(\varrho_{a})_{a\in A}$ of risk measures.
\par
The goal of risk sharing is:
\begin{equation*}
    \textrm{minimize the total risk }\int_{A}\varrho_{a}(X_{a})\mu(da)\textrm{ over all $\mathcal{X}$-feasible allocations $(X_{a})_{a\in A}$}.
\end{equation*}
A necessary and sufficient condition for the above integral to be well-defined is the measurability of $a\longmapsto\varrho_{a}(X_{a})$ (in which case integration is defined via (\ref{eq:extended-lebesgue-int})), motivating the following definition.
\begin{definition}\label{def:a-meas-pref}
    An indexed collection $(\varrho_{a})_{a\in A}$ of risk measures is said to be $\mathscr{A}$-measurable if, for each $\mathscr{A}$-measurable allocation $(X_{a})_{a\in A}$, the real-valued function $a\longmapsto\varrho_{a}(X_{a})$ is $\mathscr{A}$-measurable.
\end{definition}
For the rest of the paper, we assume $(\varrho_{a})_{a\in A}$ denotes a family satisfying Definition \ref{def:a-meas-pref}. Given any such family, an extension of the ordinary infimal convolution can be defined, representing the value function of the associated risk sharing problem.
\begin{definition}\label{def:integral-inf-conv}
    The integral infimal convolution of $(\varrho_{a})_{a\in A}$, denoted $\operatorname{\Box}_{a\in A}\varrho_{a}\mu(da)$, is defined as
    \begin{equation*}
        \left(\operatorname{\Box}_{a\in A}\varrho_{a}\mu(da)\right)(\mathcal{X})=\inf_{(X_{a})_{a\in A}\in\mathbb{A}(\mathcal{X})}\int_{A}\varrho_{a}(X_{a})\mu(da)
    \end{equation*}
    for $\mathcal{X}\in L^{\infty}(\mathbb{P})$.
\end{definition}
A set $B\in\mathscr{A}$ with $\mu(B)>0$ is called a coalition. If $B$ is a coalition, then $B$ equipped with the corresponding restriction of $\mu$ is a measure space, so the integral infimal convolution $\operatorname{\Box}_{b\in B}\varrho_{b}\mu(db)$ over $B$ is well-defined.
\par
We now briefly study some basic properties of the integral infimal convolution. The integral infimal convolution $\operatorname{\Box}_{a\in A}\varrho_{a}\mu(da)$ is said to be globally finite if $-\infty<\left(\operatorname{\Box}_{a\in A}\varrho_{a}\mu(da)\right)(\mathcal{X})<\infty$ for each $\mathcal{X}$. The integral infimal convolution $\operatorname{\Box}_{a\in A}\varrho_{a}\mu(da)$ is said to inherit a property \textit{P} from $(\varrho_{a})_{a\in A}$ if, whenever each $\varrho_{a}$ satisfies \textit{P}, $\operatorname{\Box}_{a\in A}\varrho_{a}\mu(da)$ either satisfies \textit{P} or otherwise fails global finiteness. It turns out that global finiteness only requires finiteness at a single point, and properties (\ref{it:monotonicity}), (\ref{it:cash-add}), (\ref{it:convexity}), (\ref{it:pos-hom}), and (\ref{it:cont-from-above}) are inherited by the integral infimal convolution.
\begin{proposition}\label{prop:basic-properties}
    The integral infimal convolution $\operatorname{\Box}_{a\in A}\varrho_{a}\mu(da)$ is globally finite whenever
    \begin{equation}\label{eq:finite-single-point-x0}
        -\infty<\left(\operatorname{\Box}_{a\in A}\varrho_{a}\mu(da)\right)(\mathcal{X}_{0})<\infty
    \end{equation}
    for some $\mathcal{X}_{0}\in L^{\infty}(\mathbb{P})$. The integral infimal convolution inherits the following properties from $(\varrho_{a})_{a\in A}$: being a risk measure, convexity, positive homogeneity, and continuity from above.
\end{proposition}
\begin{proof}
    If $\mathcal{X}_{0}$ is from (\ref{eq:finite-single-point-x0}) and $\mathcal{X}\in L^{\infty}(\mathbb{P})$ is arbitrary,
    \begin{equation*}
        \left(\operatorname{\Box}_{a\in A}\varrho_{a}\mu(da)\right)(\mathcal{X})=\inf_{(X_{a})_{a\in A}\in\mathbb{A}(0)}\int_{A}\varrho_{a}(X_{a}+(\mathcal{X}_{0}+\mathcal{X}-\mathcal{X}_{0})/\mu(A))
    \end{equation*}
    \begin{equation*}
        \geq\inf_{(X_{a})_{a\in A}\in\mathbb{A}(0)}\int_{A}\varrho_{a}(X_{a}+\mathcal{X}_{0}/\mu(A)-\Vert{\mathcal{X}-\mathcal{X}_{0}}\Vert_{L^{\infty}}/\mu(A))
    \end{equation*}
    \begin{equation*}
        =-\Vert{\mathcal{X}-\mathcal{X}_{0}}\Vert_{L^{\infty}}+\inf_{(X_{a})_{a\in A}\in\mathbb{A}(0)}\int_{A}\varrho_{a}(X_{a}+\mathcal{X}_{0}/\mu(A))
    \end{equation*}
    \begin{equation*}
        \geq-\Vert{\mathcal{X}-\mathcal{X}_{0}}\Vert_{L^{\infty}}+\left(\operatorname{\Box}_{a\in A}\varrho_{a}\mu(da)\right)(\mathcal{X}_{0})>-\infty.
    \end{equation*}
    A similar argument shows $\left(\operatorname{\Box}_{a\in A}\varrho_{a}\mu(da)\right)(\mathcal{X})<\infty$. Hence $-\infty<\left(\operatorname{\Box}_{a\in A}\varrho_{a}\mu(da)\right)(\mathcal{X})<\infty$ for all $\mathcal{X}$, as desired.
    \par
    The property of being a risk measure, convexity, and positive homogeneity are all obviously inherited, while a similar argument to \cite[Proposition 2]{melnikov-general-theory} shows continuity from above is inherited.
\end{proof}
\subsection{Relation to Discrete Risk Sharing}\label{subsec:discrete-risk-share-relate}
For risk measures $(\varrho_{i})_{i=1}^{N}$ and agent space $A=\{1,\dots,N\}$, the classical $N$-agent risk sharing problem takes the form
\begin{equation}\label{eq:classical-risk-sharing}
    \textrm{minimize }\sum_{i=1}^{N}\varrho_{i}(X_{i}),\textrm{ subject to }\sum_{i=1}^{N}X_{i}=\mathcal{X}.
\end{equation}
The value function of (\ref{eq:classical-risk-sharing}) is the infimal convolution $\operatorname{\Box}_{i=1}^{N}\varrho_{i}$, which coincides with the integral infimal convolution $\operatorname{\Box}_{a\in A}\varrho_{a}\mu(da)$ for the counting measure $\mu=\sum_{i=1}^{N}\delta_{i}$ (here we momentarily depart from the convention that $\mu$ denotes a non-atomic measure). Another conceivable approach, after ``normalizing'' (\ref{eq:classical-risk-sharing}), is
\begin{equation}\label{eq:normalized-risk-sharing}
    \textrm{minimize }\frac{1}{N}\sum_{i=1}^{N}\varrho_{i}(X_{i}),\textrm{ subject to }\frac{1}{N}\sum_{i=1}^{N}X_{i}=\mathcal{X}.
\end{equation}
The value function of (\ref{eq:normalized-risk-sharing}) is the infimal integral convolution $\operatorname{\Box}_{a\in A}\varrho_{a}\nu(da)$ for the uniform probability measure $\nu=\frac{1}{N}\sum_{i=1}^{N}\delta_{i}$. It is this normalized problem (\ref{eq:normalized-risk-sharing}) that is the discrete analogue of the risk sharing problem for a continuum of agents, since the agent space measure corresponding to (\ref{eq:classical-risk-sharing}) is unbounded as $N\to\infty$. Sometimes the two approaches yield the same value function, including when preferences are positively homogeneous, although even then the approaches differ in which allocations they consider admissible.
\par
When all agents have the same positively homogeneous preference, it is possible to explicitly connect discrete risk sharing with a continuum of agents (see Theorem \ref{thm:coherent}).
\section{Convexity of the Value Function for Finite Probability Spaces}\label{sec:finite-prob-space-conv}
In this section, we establish convexity of the integral infimal convolution whenever the risk measures are defined on a finite probability space.
\begin{theorem}\label{thm:finite-conv}
    Assume the following holds.
    \begin{enumerate}
        \item $\mathscr{F}$ is a finite set.
        \item The integral infimal convolution $\operatorname{\Box}_{a\in A}\varrho_{a}\mu(da)$ is globally finite.
    \end{enumerate}
    Then the integral infimal convolution $\operatorname{\Box}_{a\in A}\varrho_{a}\mu(da)$ is a convex risk measure.
\end{theorem}
The proof of Theorem \ref{thm:finite-conv} is in Appendix \ref{sec:prf-finite-conv}, and relies on the notion of Aumann integration, recalled in Appendix \ref{sec:auxiliary}.
\par
Theorem \ref{thm:finite-conv} implies the techniques of convex duality can be applied to the risk sharing problem whenever agents form a continuum, even if individual risk measures are not convex. The ability to use convex duality in the case of finite probability spaces is particularly important, since tools available in the continuous case—for example, comonotonic improvement—often assume non-atomicity of the underlying probability space.
\par
Theorem \ref{thm:finite-conv} is also important as an instrumental result. Later, in \S\ref{sec:conv-non-atom}, we state an analogue of Theorem \ref{thm:finite-conv} for non-atomic $\mathbb{P}$, and the proof is based on a finite-dimensional approximation argument relying heavily on Theorem \ref{thm:finite-conv}.
\section{Convexity of the Value Function Under Positive Homogeneity}\label{sec:coherent}
When $L^{\infty}$ is not finite-dimensional, the arguments used to prove the results of \S\ref{sec:finite-prob-space-conv} do not apply, but it is nevertheless possible to obtain similar results. For risk measures which are not positively homogeneous, this is postponed until \S\ref{sec:conv-non-atom}, but we will now see that under positive homogeneity the difficulty is greatly reduced. An important tool to that end is the notion of the minimal measure of a risk measure, introduced by Wang \cite{regsgetarbed}.
\begin{definition}
    The minimal measure $\Psi_{\varrho}$ of a positively homogeneous risk measure $\varrho$ is
    \begin{equation*}
        \Psi_{\varrho}=\inf_{N\in\mathbb{N}}\operatorname{\Box}_{i=1}^{N}\varrho.
    \end{equation*}
\end{definition}
The minimal measure was originally defined in the context of regulatory arbitrage, but admits another economic interpretation. The $N$-fold infimal convolution $\operatorname{\Box}_{i=1}^{N}\varrho$ represents the total risk after trading in an $N$-agent ``replica economy'', where new agents are added with identical preferences to the first (see \cite[Chapter 2]{core-large-econ-book}). So $\Psi_{\varrho}$ is the limit of the \textit{ex post} total risk in an $N$-agent replica economy as $N\to\infty$, connecting the minimal measure to the asymptotics of replica economies (cf. \S\ref{subsec:discrete-risk-share-relate}).
\par
The following result, establishing convexity of the minimal measure, is essentially due to Wang \cite{regsgetarbed}.
\begin{proposition}\label{prop:minimal-measure-coherent}
    Let $\varrho$ be a positively homogeneous risk measure with the Lebesgue property. If $\Psi_{\varrho}$ is globally finite, then $\Psi_{\varrho}$ is a coherent risk measure with the Lebesgue property.
\end{proposition}
\begin{proof}
    By \cite[Proposition 3.4]{regsgetarbed}, $\Psi_{\varrho}$ is a coherent risk measure. For each $N\in\mathbb{N}$, since $\varrho$ has the Lebesgue property, $\operatorname{\Box}_{i=1}^{N}\varrho$ is continuous from above, and hence $\Psi_{\varrho}=\inf_{N\in\mathbb{N}}\operatorname{\Box}_{i=1}^{N}\varrho$ is continuous from above (use the same argument as Proposition \ref{prop:checking-lebesgue-consistent}). A coherent risk measure that is continuous from above has the Lebesgue property (see \cite[Remark 4.19]{stoch-fin}), so $\Psi_{\varrho}$ has the Lebesgue property.
\end{proof}
There is a close connection between the minimal measure and the integral infimal convolution, at least when agents share the same positively homogeneous risk preference.
\begin{theorem}\label{thm:coherent}
    Suppose $\varrho$ is a positively homogeneous risk measure with the Lebesgue property, and $\varrho_{a}=\varrho$ for all $a\in A$. Then $(\varrho_{a})_{a\in A}$ is $\mathscr{A}$-measurable, and
    \begin{equation}\label{eq:inf-convolution-equivalence}
        \operatorname{\Box}_{a\in A}\varrho_{a}\mu(da)=\Psi_{\varrho}.
    \end{equation}
    In particular, if either of the terms in (\ref{eq:inf-convolution-equivalence}) is globally finite, $\operatorname{\Box}_{a\in A}\varrho_{a}\mu(da)$ is a coherent risk measure with the Lebesgue property.
\end{theorem}
The equivalence (\ref{eq:inf-convolution-equivalence}) requires positive homogeneity, as demonstrated by the entropic risk measure $\mathrm{Ent}(\mathcal{X})=\log\left(\mathbb{E}^{\mathbb{P}}\left(e^{-\mathcal{X}}\right)\right)$. Indeed, if $\varrho_{a}=\mathrm{Ent}$ for all $a$ and $\mu(A)=1$, then $\operatorname{\Box}_{a\in A}\varrho_{a}\mu(da)=\mathrm{Ent}$ (see \cite[\S4.1]{melnikov-general-theory}), but $\inf_{N\in\mathbb{N}}\operatorname{\Box}_{i=1}^{N}\mathrm{Ent}=\mathbb{E}^{\mathbb{P}}\neq\mathrm{Ent}$ (see \cite[Example 2.2]{regsgetarbed}).
\begin{proof}
    The $\mathscr{A}$-measurability of $(\varrho_{a})_{a\in A}$ follows from the Lebesgue property of $\varrho$ (see Point \ref{fact:measurability-equiv} in Appendix \ref{sec:auxiliary}).
    \par
    We first show that $\operatorname{\Box}_{a\in A}\varrho_{a}\mu(da)\leq\Psi_{\varrho}$. Fix any $N\in\mathbb{N}$ and $\mathcal{X}\in L^{\infty}(\mathbb{P})$. We may write
    \begin{equation}\label{eq:self-inf-conv-def}
        \left(\operatorname{\Box}_{i=1}^{N}\varrho\right)(\mathcal{X})=\inf_{(Z_{i})_{i=1}^{N}\in\mathbb{A}_{N}(\mathcal{X})}\sum_{i=1}^{N}\varrho(Z_{i})
    \end{equation}
    where
    \begin{equation*}
        \mathbb{A}_{N}(\mathcal{X})=\left\{(Z_{i})_{i=1}^{N}\in\left(L^{\infty}(\mathbb{P})\right)^{N}:\sum_{i=1}^{N}Z_{i}=\mathcal{X}\right\}.
    \end{equation*}
    Take any $(Z_{i})_{i=1}^{N}\in\mathbb{A}_{N}(\mathcal{X})$. Since $\mu$ is non-atomic, we may find an $\mathscr{A}$-measurable partition $(A_{i})_{i=1}^{N}$ of $A$ with $\mu(A_{i})=\mu(A)/N$ for all $i$. Define an allocation $(X_{a})_{a\in A}\in\mathbb{A}(\mathcal{X})$ by
    \begin{equation*}
        X_{a}=\frac{N}{\mu(A)}\sum_{i=1}^{N}\mathbf{1}_{A_{i}}(a)Z_{i}.
    \end{equation*}
    From positive homogeneity, we have that
    \begin{equation*}
        \left(\operatorname{\Box}_{a\in A}\varrho_{a}\mu(da)\right)(\mathcal{X})\leq\int_{A}\varrho(X_{a})\mu(da)=\sum_{i=1}^{N}\int_{A_{i}}\varrho(Z_{i}N/\mu(A))\mu(da)
    \end{equation*}
    \begin{equation*}
        =\frac{N}{\mu(A)}\sum_{i=1}^{N}\int_{A_{i}}\varrho(Z_{i})\mu(da)=\frac{N}{\mu(A)}\sum_{i=1}^{N}\mu(A_{i})\varrho(Z_{i})=\frac{N}{\mu(A)}\sum_{i=1}^{N}\frac{\mu(A)}{N}\varrho(Z_{i})=\sum_{i=1}^{N}\varrho(Z_{i}).
    \end{equation*}
    Hence $\left(\operatorname{\Box}_{a\in A}\varrho_{a}\mu(da)\right)(\mathcal{X})\leq\sum_{i=1}^{N}\varrho(Z_{i})$ for any $(Z_{i})_{i=1}^{N}\in\mathbb{A}_{N}(\mathcal{X})$, and taking the infimum over $(Z_{i})_{i=1}^{N}\in\mathbb{A}_{N}(\mathcal{X})$ conjoined with (\ref{eq:self-inf-conv-def}) implies
    \begin{equation*}
        \left(\operatorname{\Box}_{a\in A}\varrho_{a}\mu(da)\right)(\mathcal{X})\leq\left(\operatorname{\Box}_{i=1}^{N}\varrho\right)(\mathcal{X}).
    \end{equation*}
    Taking the infimum over $N\in\mathbb{N}$ gives $\left(\operatorname{\Box}_{a\in A}\varrho_{a}\mu(da)\right)(\mathcal{X})\leq\Psi_{\varrho}(\mathcal{X})$.
    \par
    We now prove the reverse inequality. If $\Psi_{\varrho}(\mathcal{X}_{0})=-\infty$ for some $\mathcal{X}_{0}\in L^{\infty}(\mathbb{P})$, then $\Psi_{\varrho}$ is identically $-\infty$ (see \cite[Proposition 2.5]{regsgetarbed}), in which case the reverse inequality is trivial. So we may suppose that $\Psi_{\varrho}$ is globally finite, in which case $\Psi_{\varrho}$ is a coherent risk measure with the Lebesgue property (see Proposition \ref{prop:minimal-measure-coherent}). A consequence is that, defining $\mathscr{Q}=\{\mathbb{Q}:\Psi_{\varrho}^{\ast}(\mathbb{Q})=0\}$, we have the dual representation $\Psi_{\varrho}=\sup_{\mathbb{Q}\in\mathscr{Q}}\mathbb{E}^{\mathbb{Q}}$.
    \par
    By Fenchel's inequality, $\varrho\geq\Psi_{\varrho}\geq\mathbb{E}^{\mathbb{Q}}$ for any $\mathbb{Q}\in\mathscr{Q}$. If $\mathcal{X}\in L^{\infty}(\mathbb{P})$ and $(X_{a})_{a\in A}\in\mathbb{A}(\mathcal{X})$, we therefore have that
    \begin{equation*}
        \int_{A}\varrho_{a}(X_{a})\mu(da)\geq\int_{A}\mathbb{E}^{\mathbb{Q}}(X_{a})\mu(da)=\mathbb{E}^{\mathbb{Q}}(\mathcal{X}).
    \end{equation*}
    Taking the infimum over $(X_{a})_{a\in A}\in\mathbb{A}(\mathcal{X})$ and noting that $\mathcal{X}\in L^{\infty}(\mathbb{P})$ was arbitrary, we obtain
    \begin{equation*}
        \operatorname{\Box}_{a\in A}\varrho_{a}\mu(da)\geq\mathbb{E}^{\mathbb{Q}}.
    \end{equation*}
    Taking the supremum over $\mathbb{Q}\in\mathscr{Q}$ yields
    \begin{equation*}
        \operatorname{\Box}_{a\in A}\varrho_{a}\mu(da)\geq\sup_{\mathbb{Q}\in\mathscr{Q}}\mathbb{E}^{\mathbb{Q}}=\Psi_{\varrho},
    \end{equation*}
    as desired.
\end{proof}
Theorem \ref{thm:coherent}, in conjunction with later results, shows that a version of the welfare theorems holds for the preference structures covered by Theorem \ref{thm:coherent} (see Corollary \ref{corr:second-welfare-positive-homogen}).
\par
As a corollary to Theorem \ref{thm:coherent}, we get the following sufficient condition for law invariance of the integral infimal convolution.
\begin{corollary}
    Suppose $\varrho$ is a positively homogeneous $\mathbb{P}$-law invariant risk measure with the Lebesgue property, and $\varrho_{a}=\varrho$ for all $a\in A$. Then, if $\mathbb{P}$ is non-atomic, $\operatorname{\Box}_{a\in A}\varrho_{a}\mu(da)$ is $\mathbb{P}$-law invariant.
\end{corollary}
\begin{proof}
    By \cite[Example 10]{law-inv-non-conv} and non-atomicity of $\mathbb{P}$, the $N$-fold infimal convolution $\operatorname{\Box}_{i=1}^{N}\varrho$ is $\mathbb{P}$-law invariant for each $N\in\mathbb{N}$. Hence the infimum $\Psi_{\varrho}=\inf_{N\in\mathbb{N}}\operatorname{\Box}_{i=1}^{N}\varrho$ is $\mathbb{P}$-law invariant. Applying (\ref{eq:inf-convolution-equivalence}) yields the claim.
\end{proof}
\section{Improperness of the Value Function}\label{sec:improper}
In this section, we show that if an agent space is non-atomic and risk preferences are far from being convex, the integral infimal convolution must be infinite. The relevance of such results to economic theory is that we later use them to show that Pareto optima cannot exist for a large class of non-convex preferences (see Corollary \ref{corr:cong-degen-pareto}).
\subsection{Conjugately Degenerate Risk Measures}
\begin{definition}\label{def:degen}
    A risk measure $\varrho$ is said to be conjugately degenerate if $\varrho^{\ast}$ is identically infinite, i.e., $\varrho^{\ast}(\mathcal{Y})=\infty$ for all $\mathcal{Y}\in L^{1}(\mathbb{P})$.
\end{definition}
The intuition behind Definition \ref{def:degen} is that convexity of a real-valued function on a finite dimensional vector space implies that the function is bounded from below by an affine function (with the necessary qualifications, a similar statement holds in infinite dimensions). If $\varrho$ is conjugately degenerate, $\varrho$ is so far from being convex that it is not possible to control $\varrho$ from below in terms of a weak-star continuous affine function. Indeed, $\varrho^{\ast}(\mathcal{Y})<\infty$ iff there is some $r\in\mathbb{R}$ so that $\varrho(\mathcal{X})\geq\mathbb{E}^{\mathbb{P}}(\mathcal{X}\mathcal{Y})-r$ for all $\mathcal{X}\in L^{\infty}(\mathbb{P})$.
\par
The integral infimal convolution of conjugately degenerate risk measures is often improper.
\begin{theorem}\label{thm:improper}
    Let $\varrho$ be a positively homogeneous risk measure which satisfies the Lebesgue property. Assume there exists $B\in\mathscr{A}$ with $\mu(B)>0$ so $\varrho_{a}=\varrho$ for all $a\in B$. Then, if $\varrho$ is conjugately degenerate,
    \begin{equation*}
        \operatorname{\Box}_{a\in A}\varrho_{a}\mu(da)=-\infty
    \end{equation*}
    whenever $\int_{A}\vert{\varrho_{a}(0)\vert}\mu(da)<\infty$
\end{theorem}
\begin{proof}
    Note that
    \begin{equation*}
        \operatorname{\Box}_{a\in A}\varrho_{a}\mu(da)\leq\int_{A\setminus B}\varrho_{a}(0)\mu(da)+\operatorname{\Box}_{b\in B}\varrho_{b}\mu(db)
    \end{equation*}
    from which it follows that $\operatorname{\Box}_{a\in A}\varrho_{a}\mu(da)$ is identically $-\infty$ whenever $\operatorname{\Box}_{b\in B}\varrho_{b}\mu(db)$ is identically $-\infty$. We henceforth may proceed with the assumption that $B=A$.
    \par
    Denote $\eta=\operatorname{\Box}_{a\in A}\varrho_{a}\mu(da)$. Suppose $\eta$ is not identically $-\infty$, in which case $\eta$ is globally finite (see Proposition \ref{prop:basic-properties}). So Theorem \ref{thm:coherent} applies to show that $\eta$ is coherent and has the Lebesgue property. In particular, there exists $\mathbb{Q}\in\mathscr{M}_{\mathbb{P}}$ with $\eta^{\ast}(\mathbb{Q})<\infty$. However,
    \begin{equation*}
        \eta^{\ast}(\mathbb{Q})=\int_{A}\varrho^{\ast}_{a}(\mathbb{Q})\mu(da)=\infty
    \end{equation*}
    by Theorem \ref{thm:conv-conj-value}, which contradicts $\eta^{\ast}(\mathbb{Q})<\infty$.
\end{proof}
\subsection{Applications to Distortion Risk Measures}
A distortion function is a non-decreasing $g:[0,1]\longrightarrow[0,1]$ with $g(0)=0$ and $g(1)=1$. For every distortion function $g$, $\widetilde{g}(u)=1-g(1-u)$ is called the dual distortion function. Given a distortion function $g$, we define the Choquet integral $I_{g}$ via
\begin{equation*}
    I_{g}(\mathcal{X})=\int_{0}^{\infty}g\left(\mathbb{P}(\{\mathcal{X}\geq t\})\right)dt+\int_{-\infty}^{0}\left(g(\mathbb{P}(\{\mathcal{X}\geq t\}))-1\right)dt,
\end{equation*}
for $\mathcal{X}\in L^{\infty}(\mathbb{P})$. Every such $I_{g}$ is a risk measure, and risk measures of this form constitute the family of so-called distortion risk measures. A distortion risk measure is necessarily positively homogeneous and law invariant (although not in general either $\mathbb{P}$-consistent or convex). If $g$ is continuous, then $I_{g}$ has the Lebesgue property; if $g$ is left-continuous, then $I_{g}$ has the Fatou property.
\par
The convex conjugate of a distortion risk measure $I_{g}$ is closely related to the anticore of the corresponding charge $\nu=g\circ\mathbb{P}$, a widely studied concept in the theory of charges (see \cite{felix-charge} and the references therein). As a consequence, we can characterize distortion risk measures which are conjugately degenerate.
\begin{proposition}\label{prop:bondareva-shapley}
    Suppose $\mathbb{P}$ is non-atomic. A distortion risk measure $\varrho=I_{g}$ is conjugately degenerate iff we do not have that $g\geq\mathrm{Id}_{[0,1]}$ (equivalently, there exists $x_{0}\in(0,1)$ with $g(x_{0})<x_{0}$).
\end{proposition}
\begin{proof}
    The claim is a consequence of the law invariant Bondareva-Shapley theorem (see \cite[Lemma 3(i)]{felix-charge}).
\end{proof}
With conjugately degenerate distortion risk measure characterized by Proposition \ref{prop:bondareva-shapley}, we can now apply Theorem \ref{thm:improper} to distortion risk measures.
\begin{proposition}\label{prop:distort-convolution}
    Suppose $g$ is a continuous distortion function, and let $\varrho_{a}=I_{g}$ for all $a\in A$. If $\mathbb{P}$ is non-atomic and we do not have that $g\geq\mathrm{Id}_{[0,1]}$, then
    \begin{equation*}
        \operatorname{\Box}_{a\in A}\varrho_{a}\mu(da)=-\infty.
    \end{equation*}
\end{proposition}
\begin{proof}
    Apply Theorem \ref{thm:improper} conjoined with Proposition \ref{prop:bondareva-shapley}.
\end{proof}
\begin{example}\label{ex:tk-improper}
    For a parameter $\gamma$, Tversky and Kahneman \cite{tk} introduced the function $w_{\mathrm{TK}}^{\gamma}$ defined by
    \begin{equation}\label{eq:tk-distortion}
        w_{\mathrm{TK}}^{\gamma}(u)=\frac{u^{\gamma}}{(u^{\gamma}+(1-u)^{\gamma})^{\frac{1}{\gamma}}}.
    \end{equation}
    Defining $v^{\gamma}_{\mathrm{TK}}=\widetilde{w_{\mathrm{TK}}^{\gamma}}$, the corresponding distortion risk measure $I_{v^{\gamma}_{\mathrm{TK}}}$ models empirical risk preferences well. It can be easily verified that $I_{v^{\gamma}_{\mathrm{TK}}}$ satisfies the criterion given in Proposition \ref{prop:bondareva-shapley} for being conjugately degenerate, so the integral infimal convolution of the distortion risk measure corresponding to $v^{\gamma}_{\mathrm{TK}}$ is identically $-\infty$.
    \qed
\end{example}
\subsection{Sufficient Conditions for Finiteness}
Given the improperness of the value function holding for certain non-convex risk measures, it is natural to ask for general conditions ensuring properness. A weak form of belief homogeneity is a sufficient condition for finiteness.
\begin{proposition}\label{prop:val-finite}
    Assume $\int_{A}\vert{\varrho_{a}(0)}\vert\mu(da)<\infty$, and there exists an $\mathscr{A}$-measurable $\mu$-integrable $\xi$ and $\mathbb{Q}\in\mathscr{M}_{\mathbb{P}}$ with
    \begin{equation*}
        \varrho^{\ast}_{a}(\mathbb{Q})\leq\xi(a)
    \end{equation*}
    for $\mu$-a.e. $a\in A$.\footnote{The use of $\xi$, rather than a condition purely on $a\longmapsto\varrho^{\ast}_{a}(\mathbb{Q})$, is to avoid measurability issues.} Then $\operatorname{\Box}_{a\in A}\varrho_{a}\mu(da)$ is globally finite.
\end{proposition}
\begin{proof}
    Let $\mathcal{X}\in L^{\infty}(\mathbb{P})$ be arbitrary. Clearly,
    \begin{equation*}
        \left(\operatorname{\Box}_{a\in A}\varrho_{a}\mu(da)\right)(\mathcal{X})\leq\Vert{\mathcal{X}}\Vert_{L^{\infty}}+\int_{A}\vert{\varrho_{a}(0)}\vert\mu(da)<\infty
    \end{equation*}
    implying $\left(\operatorname{\Box}_{a\in A}\varrho_{a}\mu(da)\right)(\mathcal{X})<\infty$.
    \par
    If $(X_{a})_{a\in A}\in\mathbb{A}(\mathcal{X})$ is arbitrary, Fenchel's inequality implies that
    \begin{equation}\label{eq:fenchel-xi-lb}
        \varrho_{a}(X_{a})\geq\mathbb{E}^{\mathbb{Q}}(X_{a})-\xi(a)
    \end{equation}
    for $\mu$-a.e. $a\in A$. Integrating (\ref{eq:fenchel-xi-lb}) and taking the infimum over $(X_{a})_{a\in A}\in\mathbb{A}(\mathcal{X})$, we get
    \begin{equation*}
        \left(\operatorname{\Box}_{a\in A}\varrho_{a}\mu(da)\right)(\mathcal{X})\geq\mathbb{E}^{\mathbb{Q}}(\mathcal{X})-\int_{A}\xi(a)\mu(da)>-\infty
    \end{equation*}
    implying $\left(\operatorname{\Box}_{a\in A}\varrho_{a}\mu(da)\right)(\mathcal{X})>-\infty$.
\end{proof}
When preferences are consistent and normalized, Proposition \ref{prop:val-finite} applies to show the integral infimal convolution is always finite.
\begin{corollary}\label{corr:val-finite-consistent}
    Suppose $\mathbb{Q}\sim\mathbb{P}$, and $\varrho_{a}$ is $\mathbb{Q}$-consistent for all $a\in A$. If $\varrho_{a}(0)=0$ for all $a\in A$, then $\operatorname{\Box}_{a\in A}\varrho_{a}\mu(da)$ is globally finite.
\end{corollary}
\begin{proof}
    For each $a\in A$ and $\mathcal{X}\in L^{\infty}(\mathbb{P})$,
    \begin{equation*}
        \varrho_{a}(\mathcal{X})\geq\varrho_{a}(\mathbb{E}^{\mathbb{Q}}(\mathcal{X}))=\varrho_{a}(0)+\mathbb{E}^{\mathbb{Q}}(\mathcal{X})=\mathbb{E}^{\mathbb{Q}}(\mathcal{X}),
    \end{equation*}
    as $\mathbb{E}^{\mathbb{Q}}(\mathcal{X})\lesssim_{c,\mathbb{Q}}\mathcal{X}$ and $\varrho_{a}$ is $\mathbb{Q}$-consistent. Hence $\varrho^{\ast}_{a}(\mathbb{Q})=0$ for all $a\in A$, and Proposition \ref{prop:val-finite} thus applies to show $\operatorname{\Box}_{a\in A}\varrho_{a}\mu(da)$ is globally finite.
\end{proof}
\section{Convexity of the Value Function for Non-Atomic Probability Spaces}\label{sec:conv-non-atom}
In this section, we study whether the integral infimal convolution is convex when the risk measures are defined on a non-atomic probability space, and are not necessarily positively homogeneous. The main result in this section, Theorem \ref{thm:general-conv}, establishes convexity and other regularity properties when preferences satisfy continuity and law-related properties. More precisely, the risk measures must have the Lebesgue property and, modulo a partition, the consistency property with respect to some probability measure—a strengthened form of law invariance with respect to that measure.
\begin{theorem}\label{thm:general-conv}
    Assume the following holds.
    \begin{enumerate}
        \item There exists a finite partition $\pi\subseteq\mathscr{A}\setminus\mu^{-1}(\{0\})$ of $A$ and a collection $(\mathbb{Q}_{B})_{B\in\pi}\sim\mathbb{P}$ such that, for each $B\in\pi$ and $a\in B$, $\varrho_{a}$ is $\mathbb{Q}_{B}$-consistent.
        \item For each $a\in A$, $\varrho_{a}$ has the Lebesgue property.
        \item The integral infimal convolution $\operatorname{\Box}_{a\in A}\varrho_{a}\mu(da)$ is globally finite.
        \item The integral $\int_{A}\vert{\varrho_{a}(0)}\vert\mu(da)$ is finite.
        \item $\mathbb{P}$ is non-atomic.
    \end{enumerate}
    Then the integral infimal convolution $\operatorname{\Box}_{a\in A}\varrho_{a}\mu(da)$ is a convex risk measure with the Lebesgue property.
\end{theorem}
The proof of Theorem \ref{thm:general-conv} is contained in Appendix \ref{sec:prf-gen-conv}. It is based on a finite dimensional approximation argument.
\par
Theorem \ref{thm:general-conv}, in conjunction with later results, shows that a version of the welfare theorems holds for consistent risk preferences (see Corollary \ref{corr:second-welfare-thm-consistent}).
\subsection{Discussion of the Assumptions}
The assumptions of Theorem \ref{thm:general-conv} are relatively weak, excepting the first. For example, the second assumption, requiring the Lebesgue property of each $\varrho_{a}$, can be checked with Proposition \ref{prop:checking-lebesgue-consistent}, while the third assumption, requiring finiteness of the integral infimal convolution, can be checked with Proposition \ref{prop:val-finite} or Corollary \ref{corr:val-finite-consistent}.
\par
The first assumption of Theorem \ref{thm:general-conv} allows for belief heterogeneity, but only within reasonable limits (the form it takes is similar to that assumed by Liebrich \cite{felixfs}). Indeed, our proof strategy relies on belief heterogeneity being limited to disagreement between finitely many coalitions of agreeing agents. In such a case, the integral infimal convolution can be decomposed into an ordinary infimal convolution $\operatorname{\Box}_{B\in\pi}\varrho_{B}$ of finitely many risk measures $(\varrho_{B})_{B\in\pi}$, where $\varrho_{B}$ is the integral infimal convolution of $(\varrho_{b})_{b\in B}$. Once this reduction is made, it is no loss of generality to assume $\pi$ is the singleton $\{A\}$, in which case agents have homogeneous beliefs. Such a reduction would not be possible if each agent $a$ were allowed to individually pick a distinct probability measure $\mathbb{Q}_{a}$.
\par
The use of the reduction to homogeneous beliefs is that, when there is agreement on a prior $\mathbb{Q}$, the infimal integral convolution is $\mathbb{Q}$-consistent. In the context of the proof of Theorem \ref{thm:general-conv}, $\mathbb{Q}$-consistency is the crucial step to ensure the infimal integral convolution has the Lebesgue property, which is necessary for the finite-dimensional approximation argument we use.
\subsection{The Role of Non-Atomicity}
Theorem \ref{thm:general-conv} relies on non-atomicity of $\mu$ to show convexity of the integral infimal convolution. The proof uses an approximation argument based on the finite-dimensional Theorem \ref{thm:finite-conv}.
\par
Theorem \ref{thm:finite-conv} is a corollary of Lyapunov-Richter convexity, which guarantees the convexity of Aumann integrals in finite dimensions (see Appendix \ref{sec:auxiliary} for definitions related to Aumann integration and for their relation to the integral infimal convolution). A natural question is whether infinite dimensional variants of Lyapunov-Richter convexity can help prove Theorem \ref{thm:general-conv}. Unfortunately, such methods in general only show that the weak-star lower semicontinuous envelope of the integral infimal convolution is convex, but this need not imply convexity—see Example \ref{ex:var-counter-integral} below.
\par
There is an exception to the impotence of infinite-dimensional arguments in the context of Theorem \ref{thm:general-conv}, at least when $\mu$ is sufficiently non-atomic. We expand on this below, with the notion of a saturated measure.
\begin{definition}
    A complete non-atomic measure space $(A,\mathscr{A},\mu)$ is said to be saturated if the following holds. For all $B\in\mathscr{A}$ with $\mu(B)>0$, the trace $\sigma$-algebra
    \begin{equation*}
        \mathscr{A}_{B}=\{C\cap B:C\in\mathscr{A}\}
    \end{equation*}
    is not countably generated modulo $\mu$-null sets.
\end{definition}
Over a saturated measure space, every Aumann integral valued in a dual Banach space with separable predual is convex (see \cite[Proposition 1]{sun-yannel}). In the finite-dimensional setting, it is exactly this kind of result that ensured the convexity of the integral infimal convolution. The exact same argument as in the finite-dimensional case therefore shows the following strengthening of Theorem \ref{thm:general-conv} for saturated measure spaces:
\begin{proposition}\label{prop:saturated}
    Assume the following holds.
    \begin{enumerate}
        \item $(A,\mathscr{A},\mu)$ is a saturated measure space.
        \item The integral infimal convolution $\operatorname{\Box}_{a\in A}\varrho_{a}\mu(da)$ is globally finite.
    \end{enumerate}
    Then the integral infimal convolution $\operatorname{\Box}_{a\in A}\varrho_{a}\mu(da)$ is a convex risk measure.
\end{proposition}
We do not otherwise pursue directions predicated on $\mu$ being saturated, since they follow easily from the existing literature while also subtracting economic content. Indeed, being non-separable, a saturated measure space can no longer be viewed as a limiting case of finite agent models (cf. \S\ref{subsec:discrete-risk-share-relate}). A survey of the applications of saturated measure spaces to optimization problems can be found in Sagara \cite{saturated-opt}.
\section{The Convex Conjugate of the Value Function}\label{sec:conv-conj-value}
In this section, we provide a computationally tractable formula for the convex conjugate of an integral infimal convolution of risk measures, even if those risk measures are not convex. In light of the convexity results of \S\ref{sec:finite-prob-space-conv} and \S\ref{sec:conv-non-atom}, guaranteeing the existence of a dual representation for the integral infimal convolution, the result is of major use, as it provides a computationally tractable formula for this dual representation.
\begin{theorem}\label{thm:conv-conj-value}
    Assume the following holds.
    \begin{enumerate}
        \item For each $a\in A$, $\varrho_{a}$ has the Lebesgue property.
        \item The integral infimal convolution $\operatorname{\Box}_{a\in A}\varrho_{a}\mu(da)$ is globally finite.
        \item The integral $\int_{A}\vert{\varrho_{a}(0)}\vert\mu(da)$ is finite.
    \end{enumerate}
    Then, for each $\mathbb{Q}\in\mathscr{M}_{\mathbb{P}}$, the map $a\longmapsto\varrho^{\ast}_{a}(\mathbb{Q})$ is $\mathscr{A}$-measurable, and
    \begin{equation*}
        \left(\operatorname{\Box}_{a\in A}\varrho_{a}\mu(da)\right)^{\ast}(\mathbb{Q})=\int_{A}\varrho^{\ast}_{a}(\mathbb{Q})\mu(da).
    \end{equation*}
\end{theorem}
The proof of Theorem \ref{thm:conv-conj-value} is in Appendix \ref{sec:prf-formula-conj}. Unlike most of the major results of this paper, we do not invoke non-atomicity of $\mu$ in the proof.
\par
If $(\varrho_{i})_{i=1}^{N}$ is a finite sequence of risk measures, we have that
\begin{equation}\label{eq:finite-conj-value-formula}
    \left(\operatorname{\Box}_{i=1}^{N}\varrho_{i}\right)^{\ast}=\sum_{i=1}^{N}\varrho^{\ast}_{i}.
\end{equation}
The formula (\ref{eq:finite-conj-value-formula}) has been generalized to other contexts, including countably many agents (see Righi and Moresco \cite{righi-op}) or a measure space of agents with convex preferences (see \cite{melnikov-general-theory}); as a substitute for the right side of (\ref{eq:finite-conj-value-formula}), the former generalization uses a weighted sum, while the latter generalization uses an integral. Theorem \ref{thm:conv-conj-value} thus fits naturally within a stream of results representing the convex conjugate of the risk sharing value function via summation or integration.
\par
When the integral infimal convolution is convex and satisfies the preconditions of Theorem \ref{thm:conv-conj-value}, Theorem \ref{thm:conv-conj-value} provides a duality-based formula for the integral infimal convolution as
\begin{equation*}
    \operatorname{\Box}_{a\in A}\varrho_{a}\mu(da)=\sup_{\mathbb{Q}\in\mathscr{M}_{\mathbb{P}}}\left(\mathbb{E}^{\mathbb{Q}}-\int_{A}\varrho^{\ast}_{a}(\mathbb{Q})\mu(da)\right).
\end{equation*}
\section{Applications to Pareto Efficiency and Competitive Equilibria}\label{sec:pareto-equi}
In economic theory, when preferences are convex, there is a relation between Pareto efficiency, minimizers of a social loss function, and the existence of competitive equilibria. We establish these relations for non-convex preferences, which present distinct problems from the classical convex case.
\subsection{Pareto Efficiency}
\begin{definition}\label{def:pareto-efficient}
    Let $(\varrho_{a})_{a\in A}$ be an $\mathscr{A}$-measurable collection of preferences. An $\mathcal{X}$-feasible allocation $(X_{a})_{a\in A}$ is said to be Pareto efficient if there does not exist another $\mathcal{X}$-feasible allocation $(Y_{a})_{a\in A}$ with $\varrho_{a}(Y_{a})\leq\varrho_{a}(X_{a})$ $\mu$-a.e. and $\mu\left(\left\{\varrho_{a}(Y_{a})<\varrho_{a}(X_{a})\right\}\right)>0$. If such a $(Y_{a})_{a\in A}$ exists for an $\mathcal{X}$-feasible $(X_{a})_{a\in A}$, $(Y_{a})_{a\in A}$ is said to be a Pareto improvement on $(X_{a})_{a\in A}$.
\end{definition}
Pareto efficiency is a standard concept in the theory of risk sharing with finitely many agents (see, for example, \cite{ghossoubzhu2025,quantile-share}). Definition \ref{def:pareto-efficient} differs from the finite agent context in that it ignores $\mu$-null sets, although otherwise the two contexts are conceptually the same. The idea behind Pareto efficiency of an allocation $(X_{a})_{a\in A}$ is that $(X_{a})_{a\in A}$ is Pareto efficient iff, for any coalition $B$, it is not possible to strictly improve the lot of all agents in $B$ without making some coalition in $A\setminus B$ worse off.
\par
As in the finite agent case, Pareto efficiency of $(X_{a})_{a\in A}$ is equivalent to $(X_{a})_{a\in A}$ being a minimizer to an infimal convolution problem, at least after supposing this problem has a finite value.
\begin{proposition}\label{prop:pareto-char}
    Suppose $(\varrho_{a})_{a\in A}$ is an $\mathscr{A}$-measurable collection of risk measures. If $\left(\operatorname{\Box}_{a\in A}\varrho_{a}\mu(da)\right)(\mathcal{X})<\infty$, then an $\mathcal{X}$-feasible allocation is Pareto efficient if, and only if,
    \begin{equation*}
        \left(\operatorname{\Box}_{a\in A}\varrho_{a}\mu(da)\right)(\mathcal{X})=\int_{A}\varrho_{a}(X_{a})\mu(da).
    \end{equation*}
    Furthermore, if $\left(\operatorname{\Box}_{a\in A}\varrho_{a}\mu(da)\right)(\mathcal{X})=-\infty$, and $\int_{A}\varrho_{a}(X_{a})\mu(da)>-\infty$, then $(X_{a})_{a\in A}$ cannot be Pareto efficient.
\end{proposition}
\begin{proof}
    The argument given in the proof of (Proposition 4, \cite{melnikov-general-theory}) shows that, for an allocation $(X_{a})_{a\in A}\in\mathbb{A}(\mathcal{X})$, the existence of $(Y_{a})_{a\in A}\in\mathbb{A}(\mathcal{X})$ with $\int_{A}\varrho_{a}(Y_{a})\mu(da)<\int_{A}\varrho_{a}(X_{a})\mu(da)$ implies the existence of a Pareto improvement on $(X_{a})_{a\in A}$. This argument yields both the first and second claims.
\end{proof}
Cash additivity is crucial for the validity of Proposition \ref{prop:pareto-char}. Indeed, the equivalence in the finite agent setting between Pareto optimality and being an optimizer of a social welfare function is, in the absence of cash additivity, predicated on convexity of individual preferences (see \cite[Proposition 2.5]{ghossoubzhu2025}). Hence there is no clear reason for a relevantly similar analogue of Proposition \ref{prop:pareto-char} to hold in the absence of both cash additivity and convexity.
\subsection{Non-existence of Pareto Optima}
The second assertion of Proposition \ref{prop:pareto-char} connects the existence of Pareto improvements to improperness of the integral infimal convolution. A consequence, as we show in this subsection, is that Pareto efficient allocations cannot exist for a wide class of preferences. In particular, Pareto efficient allocations cannot exist for a large family of non-convex distortion risk measures (see Corollary \ref{corr:cong-degen-pareto}).
\par
We deduce non-existence results for specific classes of preferences from an abstract non-existence result, Theorem \ref{thm:no-pareto-ill-posed}, which uses the notion of hereditary ill-posedness.
\begin{definition}\label{def:hereditary-not-posed}
    Let $(\varrho_{a})_{a\in A}$ be an $\mathscr{A}$-measurable collection of risk measures, and let $C\in\mathscr{A}$ have non-zero measure $\mu(C)>0$. We say the risk sharing problem for $(\varrho_{a})_{a\in A}$ is hereditarily ill-posed on $C$ if, for any $B\in\mathscr{A}$ with $\mu(B)>0$ and $B\subseteq C$,
    \begin{equation*}
        \operatorname{\Box}_{b\in B}\varrho_{b}\mu(db)=-\infty.
    \end{equation*}
\end{definition}
Under hereditary ill-posedness, Pareto optima cannot exist.
\begin{theorem}\label{thm:no-pareto-ill-posed}
    Suppose $(\varrho_{a})_{a\in A}$ is an $\mathscr{A}$-measurable collection of risk measures such that the risk sharing problem for $(\varrho_{a})_{a\in A}$ is hereditarily ill-posed on $C$ for some $C\in\mathscr{A}$ with $\mu(C)>0$. Then, for any $\mathcal{X}\in L^{\infty}(\mathbb{P})$ and $(X_{a})_{a\in A}\in\mathbb{A}(\mathcal{X})$, $(X_{a})_{a\in A}$ is not Pareto efficient.
\end{theorem}
\begin{proof}
    Let $(X_{a})_{a\in A}\in\mathbb{A}(\mathcal{X})$ be arbitrary. There exists a set $B\in\mathscr{A}$ with $\mu(B)>0$ and $B\subseteq C$ so that $\int_{B}\vert{\varrho_{b}(0)}\vert\vee\Vert{X_{b}}\Vert_{L^{\infty}}\mu(db)<\infty$. We have that
    \begin{equation*}
        \left(\operatorname{\Box}_{b\in B}\varrho_{b}\mu(db)\right)\left(\int_{B}X_{b}\mu(db)\right)=-\infty,
    \end{equation*}
    as the risk sharing problem for $(\varrho_{a})_{a\in A}$ is hereditarily ill-posed on $C$, but
    \begin{equation*}
        \int_{B}\varrho_{b}(X_{b})\mu(db)\geq\int_{B}\varrho_{b}\left(-\Vert{X_{b}}\Vert_{L^{\infty}}\right)\mu(db)\geq-2\int_{B}\vert{\varrho_{b}(0)}\vert\vee\Vert{X_{b}}\Vert_{L^{\infty}}\mu(db)>-\infty
    \end{equation*}
     Hence
     \begin{equation*}
         \int_{B}\varrho_{b}(X_{b})\mu(db)>\left(\operatorname{\Box}_{b\in B}\varrho_{b}\mu(db)\right)\left(\int_{B}X_{b}\mu(db)\right).
     \end{equation*}
     By Proposition \ref{prop:pareto-char}, there exists a Pareto improvement $(Y_{b})_{b\in B}$ of $(X_{b})_{b\in B}$. Defining $(Z_{a})_{a\in A}$ by setting $Z_{a}=Y_{a}$ for $a\in B$, and setting $Z_{a}=X_{a}$ for $a\in A\setminus B$, we have that $(Z_{a})_{a\in A}$ is a Pareto improvement on $(X_{a})_{a\in A}$. So, $(X_{a})_{a\in A}$ is not Pareto efficient.
\end{proof}
The hereditary ill-posedness assumption in Theorem \ref{thm:no-pareto-ill-posed} has the advantage of often being easy to verify.
\begin{corollary}\label{corr:cong-degen-pareto}
    Suppose $g$ is a continuous distortion function, and let $\varrho_{a}=I_{g}$ for all $a\in A$. If $\mathbb{P}$ is non-atomic and we do not have that $g\geq\mathrm{Id}_{[0,1]}$, then for any $\mathcal{X}\in L^{\infty}(\mathbb{P})$ and $(X_{a})_{a\in A}\in\mathbb{A}(\mathcal{X})$, $(X_{a})_{a\in A}$ is not Pareto efficient.
\end{corollary}
\begin{proof}
    The result is a joint consequence of Theorem \ref{thm:no-pareto-ill-posed} and Proposition \ref{prop:distort-convolution}.
\end{proof}
\begin{example}
    Corollary \ref{corr:cong-degen-pareto} applies to the empirically realistic risk preferences discussed in Example \ref{ex:tk-improper}. Hence, if individual preferences are given by $\varrho_{a}=I_{v_{\mathrm{TK}}^{\gamma}}$, no Pareto optima exist.
    \qed
\end{example}
Hereditary ill-posedness implies that the integral infimal convolution is identically $-\infty$. There are general conditions under which the integral infimal convolution is finite (see Proposition \ref{prop:val-finite}), so Theorem \ref{thm:no-pareto-ill-posed} does not by itself rule out Pareto optima entirely. Indeed, we now give two examples to show that neither improperness of the integral infimal convolution nor non-convexity of preferences implies the non-existence of Pareto optima. In particular, hereditary ill-posedness cannot be weakened to improperness in the context of Theorem \ref{thm:no-pareto-ill-posed}.
\begin{example}
    Fix some $\mathscr{A}$-measurable $\xi$ with $\int_{A}\xi(a)\mu(da)=-\infty$, and let $\varrho_{a}=\mathbb{E}^{\mathbb{P}}+\xi(a)$ for each $a\in A$. Then $\operatorname{\Box}_{a\in A}\varrho_{a}=-\infty$, but any $\mathcal{X}$-feasible allocation is Pareto efficient.
    \qed
\end{example}
\begin{example}\label{ex:PO-exist-non-convex}
    Let $g$ and $h$ be concave distortion functions such that $g<h$ on $(0,u_{0})$ and $h<g$ on $(u_{0},1)$ for some $u_{0}\in(0,1)$, and let $\varrho=I_{g}\wedge I_{h}$ (note that $\varrho$ is not convex if $\mathbb{P}$ is non-atomic). Suppose individual preferences are given by $\varrho_{a}=\varrho$ for all $a\in A$. Fix a bounded random variable $\mathcal{X}\geq0$ whose law is non-atomic, find $x_{0}\geq0$ so that $\mathbb{P}(\{\mathcal{X}\geq x_{0}\})=u_{0}$, and split $A$ into two disjoint sets $A_{1}$ and $A_{2}$ of equal measure. Then a Pareto optimal allocation $(X^{\ast}_{a})_{a\in A}\in\mathbb{A}(\mathcal{X})$ is given by
    \begin{equation*}
        X^{\ast}_{a}=
        \begin{cases}
            \frac{2}{\mu(A)}\left(\mathcal{X}\wedge x_{0}\right),\textrm{ if }a\in A_{1};
            \\
            \frac{2}{\mu(A)}\left(\mathcal{X}\vee x_{0}-x_{0}\right),\textrm{ if }a\in A_{2}.
        \end{cases}
    \end{equation*}
    Indeed, $\operatorname{\Box}_{a\in A}\varrho_{a}\mu(da)=I_{g\wedge h}$ (use Theorem \ref{thm:coherent} conjoined with Theorem \ref{thm:conv-conj-value}), and a direct calculation yields
    \begin{equation*}
        \int_{A}\varrho_{a}(X^{\ast}_{a})\mu(da)=\varrho\left(\mathcal{X}\wedge x_{0}\right)+\varrho\left(\mathcal{X}\vee x_{0}-x_{0}\right)\leq I_{h}\left(\mathcal{X}\wedge x_{0}\right)+I_{g}\left(\mathcal{X}\vee x_{0}-x_{0}\right)
    \end{equation*}
    \begin{equation*}
        =I_{g\wedge h}\left(\mathcal{X}\wedge x_{0}\right)+I_{g\wedge h}\left(\mathcal{X}\vee x_{0}-x_{0}\right)=I_{g\wedge h}(\mathcal{X})
    \end{equation*}
    by comonotonic additivity.
    \par
    An example of $g$, $h$, $u_{0}$, $\mathcal{X}$, and $x_{0}$ as above is $g(u)=u^{1/2}$, $h(u)=\frac{1}{2}u^{1/4}+\frac{1}{2}u$, $u_{0}\approx0.145898$, $\mathcal{X}\sim\mathrm{Unif}[0,1]$, and $x_{0}=1-u_{0}\approx 0.854102$.
    \qed
\end{example}
\subsection{Competitive Equilibria}
Competitive equilibria are defined for economies, which consist of an aggregate endowment, an initial endowment, and agent preferences.
\begin{definition}
    An economy is a $3$-tuple $\mathscr{E}=(\mathcal{X},(E_{a})_{a\in A},(\varrho_{a})_{a\in A})$, where $\mathcal{X}\in L^{\infty}(\mathbb{P})$ is an aggregate endowment, $(E_{a})_{a\in A}\in\mathbb{A}(\mathcal{X})$ is an initial endowment, and $(\varrho_{a})_{a\in A}$ is an $\mathscr{A}$-measurable collection of risk measures, representing individual preferences.
\end{definition}
\begin{definition}\label{def:comp-equi}
    For an economy $\mathscr{E}=(\mathcal{X},(E_{a})_{a\in A},(\varrho_{a})_{a\in A})$, a tuple $(\mathbb{Q},(X_{a})_{a\in A})\in\mathscr{M}_{\mathbb{P}}\times\mathbb{A}(\mathcal{X})$ is said to be an $\mathscr{E}$-competitive equilibrium if, for $\mu$-a.e. $a\in A$, $X_{a}$ solves the problem
    \begin{equation*}
        \varrho_{a}(Y)\longrightarrow\mathrm{min!}
    \end{equation*}
    subject to the constraint $\mathbb{E}^{\mathbb{Q}}(E_{a})\leq\mathbb{E}^{\mathbb{Q}}(Y)$.
\end{definition}
\subsection{The Welfare Theorems}
Pareto efficiency is related to equilibrium, a relationship summed up by the two welfare theorems. The first welfare theorem states that an equilibrium is Pareto efficient. The second welfare theorem states that a Pareto optimum can be turned into an equilibrium after making transfer payments. Unfortunately, the second welfare theorem relies on convexity of individual preferences. However, it is known that ``large'' economies sometimes admit versions of the second welfare theorem without the assumption of individual preference convexity (see Anderson \cite{welfare-theorem-second-anderson}).
\par
We find that convexity of the integral infimal convolution, but not necessarily of individual preferences, is a sufficient condition for the welfare theorems to hold.
\begin{theorem}\label{thm:welfare-theorems}
    For any economy $\mathscr{E}=(\mathcal{X},(E_{a})_{a\in A},(\varrho_{a})_{a\in A})$, we have the following.
    \begin{enumerate}
        \item If $(\mathbb{Q},(X_{a})_{a\in A})$ is an $\mathscr{E}$-competitive equilibrium, then $(X_{a})_{a\in A}$ is Pareto efficient.
        \item Suppose the infimal integral convolution $\operatorname{\Box}_{a\in A}\varrho_{a}\mu(da)$ is convex and proper, and each $\varrho_{a}$ has the Lebesgue property. If $(X_{a})_{a\in A}$ is Pareto efficient, then there exists a measure $\mathbb{Q}\in\mathscr{M}_{\mathbb{P}}$ so that, for $(Y_{a})_{a\in A}$ defined as $Y_{a}=X_{a}+\mathbb{E}^{\mathbb{Q}}(E_{a}-X_{a})$, $(\mathbb{Q},(Y_{a})_{a\in A})$ is an $\mathscr{E}$-competitive equilibrium.
    \end{enumerate}
\end{theorem}
The proof of Theorem \ref{thm:welfare-theorems} does not invoke non-atomicity of $\mu$. The largeness of an economy is useful only insofar as it guarantees convexity of the integral infimal convolution.
\begin{proof}
    We prove the first assertion here; the proof of the second assertion is contained in Appendix \ref{sec:prf-welfare}.
    \par
    Let $(\mathbb{Q},(X_{a})_{a\in A})$ be an $\mathscr{E}$-competitive equilibrium. Suppose, for the sake of contradiction, that there exists a Pareto improvement $(Y_{a})_{a\in A}$ on $(X_{a})_{a\in A}$. In particular, $\varrho_{a}(Y_{a})\leq\varrho_{a}(X_{a})$ for $\mu$-a.e. $a\in A$, with strict inequality on a set of non-zero $\mu$-measure. Hence $\mathbb{E}^{\mathbb{Q}}(Y_{a})\leq\mathbb{E}^{\mathbb{Q}}(E_{a})$ for $\mu$-a.e. $a\in A$, with strict inequality on a set of non-zero $\mu$-measure. Integrating, we obtain
    \begin{equation*}
        \mathbb{E}^{\mathbb{Q}}(\mathcal{X})=\int_{A}\mathbb{E}^{\mathbb{Q}}(Y_{a})\mu(da)<\int_{A}\mathbb{E}^{\mathbb{Q}}(E_{a})\mu(da)=\mathbb{E}^{\mathbb{Q}}(\mathcal{X}),
    \end{equation*}
    a contradiction.
\end{proof}
We list some corollaries to Theorem \ref{thm:welfare-theorems}, which establish versions of the second welfare theorem for specific classes of non-convex preferences.
\begin{corollary}\label{corr:second-welfare-thm-consistent}
    Assume the following holds for an $\mathscr{A}$-measurable preference collection $(\varrho_{a})_{a\in A}$.
    \begin{enumerate}
        \item There exists a finite partition $\pi\subseteq\mathscr{A}\setminus\mu^{-1}(\{0\})$ of $A$ and a collection $(\mathbb{Q}_{B})_{B\in\pi}\sim\mathbb{P}$ such that, for each $B\in\pi$ and $a\in B$, $\varrho_{a}$ is $\mathbb{Q}_{B}$-consistent.
        \item For each $a\in A$, $\varrho_{a}$ has the Lebesgue property.
        \item The integral infimal convolution $\operatorname{\Box}_{a\in A}\varrho_{a}\mu(da)$ is globally finite.
        \item The function $a\longmapsto\varrho_{a}(0)$ is $\mu$-integrable.
        \item $\mathbb{P}$ is non-atomic.
    \end{enumerate}
    For any economy of the form $\mathscr{E}=(\mathcal{X},(E_{a})_{a\in A},(\varrho_{a})_{a\in A})$, if $(X_{a})_{a\in A}$ is Pareto efficient, then there exists a measure $\mathbb{Q}\in\mathscr{M}_{\mathbb{P}}$ so that, for $(Y_{a})_{a\in A}$ defined as $Y_{a}=X_{a}+\mathbb{E}^{\mathbb{Q}}(E_{a}-X_{a})$, $(\mathbb{Q},(Y_{a})_{a\in A})$ is an $\mathscr{E}$-competitive equilibrium.
\end{corollary}
\begin{proof}
    Apply Theorem \ref{thm:general-conv} to deduce that the preconditions of Theorem \ref{thm:welfare-theorems} hold. Applying Theorem \ref{thm:welfare-theorems} yields the claim.
\end{proof}
\begin{corollary}\label{corr:second-welfare-positive-homogen}
    Assume the following holds for an $\mathscr{A}$-measurable preference collection $(\varrho_{a})_{a\in A}$.
    \begin{enumerate}
        \item There exists a positively homogenous risk measure $\varrho$ with the Lebesgue property such that $\varrho_{a}=\varrho$ for all $a\in A$.
        \item The integral infimal convolution $\operatorname{\Box}_{a\in A}\varrho_{a}\mu(da)$ is globally finite.
    \end{enumerate}
    For any economy of the form $\mathscr{E}=(\mathcal{X},(E_{a})_{a\in A},(\varrho_{a})_{a\in A})$, if $(X_{a})_{a\in A}$ is Pareto efficient, then there exists a measure $\mathbb{Q}\in\mathscr{M}_{\mathbb{P}}$ so that, for $(Y_{a})_{a\in A}$ defined as $Y_{a}=X_{a}+\mathbb{E}^{\mathbb{Q}}(E_{a}-X_{a})$, $(\mathbb{Q},(Y_{a})_{a\in A})$ is an $\mathscr{E}$-competitive equilibrium.
\end{corollary}
\begin{proof}
    Apply Theorem \ref{thm:coherent} to deduce that the preconditions of Theorem \ref{thm:welfare-theorems} hold. Applying Theorem \ref{thm:welfare-theorems} yields the claim.
\end{proof}
\begin{example}
    Let $\mathscr{E}=(\mathcal{X},(E_{a})_{a\in A},(\varrho_{a})_{a\in A})$ be an economy, where $(\varrho_{a})_{a\in A}$ and $\mathcal{X}$ are the same as in Example \ref{ex:PO-exist-non-convex}, in which case the preconditions to Corollary \ref{corr:second-welfare-positive-homogen} hold. Hence the Pareto optimum $(X^{\ast}_{a})_{a\in A}$ constructed in Example \ref{ex:PO-exist-non-convex} can be turned into a competitive equilibrium after making transfer payments.
    \qed
\end{example}
\appendix
\section{Some Auxiliary Results and Definitions}\label{sec:auxiliary}
\subsection{Correspondences and Aumann Integration}\label{subsec:aumann-recall}
In this subsection, we recall the Aumann integral, which requires first recalling the notion of an integrable selector.
\begin{definition}\label{def:aumann-and-integrable-selector}
    Given a correspondence $F:A\longrightarrow 2^{L^{\infty}(\mathbb{P})}$, an integrable selector of $F$ is an $\mathscr{A}$-measurable Gelfand integrable function $(X_{a})_{a\in A}\in\left(L^{\infty}(\mathbb{P})\right)^{A}$ such that $X_{a}\in F(a)$ for $\mu$-a.e. $a\in A$. The set of all integrable selectors of $F$ is denoted $S^{1}(F)$.
    \par
    The Aumann integral $\int_{A}F(a)\mu(da)$ of $F$ is
    \begin{equation*}
        \int_{A}F(a)\mu(da)=\left\{\int_{A}X_{a}\mu(da):(X_{a})_{a\in A}\in S^{1}(F)\right\}.
    \end{equation*}
\end{definition}
\subsection{Characterization of Acceptance Sets}
An application of Aumann integration is the following characterization of the acceptance set of an integral infimal convolution, which can be viewed as a continuous variant of the characterization of the acceptance set of an infimal convolution in terms of Minkowski sums (see the proof of \cite[Theorem 4.1]{felixfs}).
\begin{theorem}\label{thm:char-acceptance}
    Suppose $\operatorname{\Box}_{a\in A}\varrho_{a}\mu(da)$ is globally finite. Then, the acceptance set $\mathfrak{A}\left(\operatorname{\Box}_{a\in A}\varrho_{a}\mu(da)\right)$ of $\operatorname{\Box}_{a\in A}\varrho_{a}\mu(da)$ is the $L^{\infty}(\mathbb{P})$-closure of the Aumann integral $\int_{A}\mathfrak{A}(\varrho_{a})\mu(da)$.
\end{theorem}
\begin{proof}
    Although stated for convex risk measures, the same proof as given for \cite[Theorem 9]{melnikov-general-theory} applies in the absence of convexity.
\end{proof}
\subsection{Measurability Notions for Correspondences: Abstract Results}\label{subsec:abstract-corr}
In the sequel, it will be useful to make measurability assumptions when we use correspondences. In this subsection, we consider two such notions, $\mathscr{A}$-measurable correspondences and correspondences with a measurable graph. Although later we only consider correspondences valued in $L^{\infty}(\mathbb{P})$, for ease of presentation in this subsection we consider correspondences valued in general classes of topological spaces.
\par
Let $(T,\mathscr{T})$ be a Hausdorff topological space with Borel $\sigma$-algebra $\mathscr{B}_{T}$. Just as a function from $A$ to $T$ is $(\mathscr{A},\mathscr{B}_{T})$-measurable if all inverse images of closed sets are $\mathscr{A}$-measurable, a similar stipulation can be made for correspondences. Given a correspondence $F:A\longrightarrow 2^{T}$, define the (lower) inverse image $F^{-1}(U)$ of a set $U\subseteq T$ to be
\begin{equation*}
    F^{-1}(U)=\{a\in A:F(a)\cap U\neq\emptyset\}.
\end{equation*}
The $\mathscr{A}$-measurability of a correspondence consists in all inverse images of closed sets being measurable.
\begin{definition}\label{def:strong-measurable-corr}
    $F:A\longrightarrow 2^{T}$ is said to be $\mathscr{A}$-measurable if $F^{-1}(U)\in\mathscr{A}$ for all $\mathscr{T}$-closed sets $U\subseteq T$.
\end{definition}
Another measurability notion is based on the graph $\Gamma(F)$ of the correspondence $F$, defined as
\begin{equation*}
    \Gamma(F)=\{(a,t)\in A\times T:t\in F(a)\}.
\end{equation*}
More precisely, we say that $F$ has a measurable graph if $\Gamma(F)\in\mathscr{A}\otimes\mathscr{B}_{T}$, where $\mathscr{B}_{T}$ denotes the Borel $\sigma$-algebra of $(T,\mathscr{T})$.
\par
Measurability of the graph is generally a stronger notion than $\mathscr{A}$-measurability. Indeed, we have the following lemma, which is a straightforward consequence of the measurable projection theorem \cite[Theorem 18.25]{aliprantis-inf-dim} and completeness of $\mu$.
\begin{lemma}\label{lem:relation-bw-two-measurability-notions}
    Suppose $(T,\mathscr{T})$ is a separable metrizable space. If $F:A\longrightarrow 2^{T}$ has a measurable graph, then $F$ is $\mathscr{A}$-measurable.
\end{lemma}
The primary reason we use correspondences with a measurable graph is that they readily admit measurable selectors, as demonstrated by the following result of Leese \cite{leese}.
\begin{proposition}\label{prop:measurable-selection-souslin}
    Suppose $F:A\longrightarrow 2^{T}$ is a correspondence with non-empty values and a measurable graph, and $(T,\mathscr{T})$ is a Souslin space (i.e., the continuous image of a Polish space). Then there exists a sequence $(f_{n})_{n=1}^{\infty}$ of $(\mathscr{A},\mathscr{B}_{T})$-measurable selectors of $F$ with $\{f_{n}(a):n\in\mathbb{N}\}$ $\mathscr{T}$-dense in $F(a)$ for each $a\in A$. In particular, an $(\mathscr{A},\mathscr{B}_{T})$-measurable selector of $F$ exists.
\end{proposition}
In the above, a selector of $F$ is a function $f:A\longrightarrow T$ such that $f(a)\in F(a)$ for all $a\in A$ (cf. Definition \ref{def:aumann-and-integrable-selector}).
\begin{proof}
    This is a joint consequence of Lemma 3 and Theorem 7 of \cite{leese}.\footnote{Remark that the results of Leese's paper assume the measurable space $(A,\mathscr{A})$ is closed under the Souslin operation—a condition from descriptive set theory which is satisfied because we assume that $\mathscr{A}$ is $\mu$-complete (see \cite[Theorem 29.16]{kechris}).}
\end{proof}
\par
The advantage of Proposition \ref{prop:measurable-selection-souslin} in our context is that it is valid for Souslin spaces. The topological spaces we later consider are Souslin spaces but not necessarily Polish spaces (see Point \ref{fact:souslin-space} below).
\par
To deduce the existence of measurable selectors, it therefore suffices to show that the correspondence has a measurable graph. We now move onto answering this second question for particular classes of correspondences. In applications, a commonly encountered class of correspondences are those of the form
\begin{equation}\label{eq:phi_u}
    a\longmapsto\left\{t\in T:\Phi(a,t)\in U\right\}
\end{equation}
for some function $\Phi:A\times T\longrightarrow V$, where $V$ is some topological space and $U\subseteq V$ is some Borel subset. If $\Phi$ is a Carathéodory function, in the sense that $\Phi(a,\cdot)$ is $\mathscr{T}$-continuous and $\Phi(\cdot,t)$ is $\mathscr{A}$-measurable for each fixed $a\in A$ and $t\in T$, then much is known about the regularity properties of (\ref{eq:phi_u}) (see \cite[Lemma 18.7]{aliprantis-inf-dim}).
\par
However, in the particular applications that arise in this paper, we may only know that $\Phi(a,\cdot)$ is continuous when restricted to an increasing sequence of closed sets covering $T$, but not on $T$ itself (cf. the Lebesgue property (\ref{it:lebesgue})). For that reason, we introduce the notion of a $\sigma$-Carathéodory function, which generalizes the concept of a Carathéodory function.
\begin{definition}\label{def:sigma-cara}
    Let $(V,\mathscr{V})$ be a topological space with Borel $\sigma$-algebra $\mathscr{B}_{V}$, and let $\Phi:A\times T\longrightarrow V$ be a function. We say $\Phi$ is a $\sigma$-Carathéodory function if there exists a sequence $(T_{n})_{n=1}^{\infty}\subseteq 2^{T}$ of $\mathscr{T}$-closed sets such that the following holds.
        \begin{enumerate}
            \item The sequence $(T_{n})_{n=1}^{\infty}$ covers $T$, i.e., $T=\bigcup_{n=1}^{\infty}T_{n}$.
            \item For each $n\in\mathbb{N}$, $\Phi|_{A\times T_{n}}$ is a Carathéodory function. That is, for each $t\in T_{n}$ and $a\in A$,
            \begin{equation*}
                T_{n}\ni t'\longmapsto\Phi(a,t')\textrm{ is continuous, }A\ni a'\longmapsto\Phi(a',t)\textrm{ is $(\mathscr{A},\mathscr{B}_{V})$-measurable}.
            \end{equation*}
        \end{enumerate}
\end{definition}
If $\Phi$ is a $\sigma$-Carathéodory function, then the correspondence (\ref{eq:phi_u}) has a measurable graph, making Proposition \ref{prop:measurable-selection-souslin} applicable to such correspondences.
\begin{proposition}\label{prop:sigma-cara-corr}
    Let $(V,\mathscr{V})$ be a separable metrizable space with Borel $\sigma$-algebra $\mathscr{B}_{V}$, and let $\Phi:A\times T\longrightarrow V$ be a function. Suppose the following holds:
    \begin{enumerate}
        \item $T$ is a separable metrizable space.
        \item $\Phi:S\times T\longrightarrow V$ is a $\sigma$-Carathéodory function.
    \end{enumerate}
    For $U\subseteq V$, define $\Phi_{U}:A\longrightarrow 2^{T}$ by
    \begin{equation*}
        \forall a\in A:\Phi_{U}(a)=\left\{t\in T:\Phi(a,t)\in U\right\}.
    \end{equation*}
    If $U$ is a Borel set in $V$, then $\Gamma(\Phi_{U})\in\mathscr{A}\otimes\mathscr{B}_{T}$ and $\Phi_{U}$ is $\mathscr{A}$-measurable.
\end{proposition}
\begin{proof}
    Let $(T_{n})_{n=1}^{\infty}$ be as in Definition \ref{def:sigma-cara}. Fix an arbitrary $n\in\mathbb{N}$, and denote by $\Psi_{n}$ the restriction $\Phi|_{A\times T_{n}}$. Being a Carathéodory function, $\Psi_{n}$ is $\mathscr{A}\otimes\mathscr{B}_{T_{n}}$-measurable (apply \cite[Lemma 4.51]{aliprantis-inf-dim}, noting that $T_{n}$ and $V$ are separable and metrizable), where $\mathscr{B}_{T_{n}}$ is the Borel $\sigma$-algebra of $T_{n}$. In particular, as $U$ is Borel in $V$,
    \begin{equation*}
        \Psi_{n}^{-1}(U)\in\mathscr{A}\otimes\mathscr{B}_{T_{n}}\subseteq\mathscr{A}\otimes\mathscr{B}_{T},
    \end{equation*}
    where the latter inclusion follows from the closedness of $T_{n}$. We have that
    \begin{equation*}
        \Gamma\left(\Phi_{U}\cap T_{n}\right)=\left\{(a,t)\in A\times T_{n}:\Phi(a,t)\in U\right\}=\Psi^{-1}_{n}(U)\in\mathscr{A}\otimes\mathscr{B}_{T},
    \end{equation*}
    where $\Gamma$ denotes the graph of a correspondence. Hence $\Gamma(\Phi_{U}\cap T_{n})\in\mathscr{A}\otimes\mathscr{B}_{T}$.
    \par
    We now let $n$ vary. We have that
    \begin{equation*}
        \Gamma(\Phi_{U})=\Gamma\left(\bigcup_{n=1}^{\infty}\left(\Phi_{U}\cap T_{n}\right)\right)=\bigcup_{n=1}^{\infty}\Gamma\left(\Phi_{U}\cap T_{n}\right)\in\mathscr{A}\otimes\mathscr{B}_{T}
    \end{equation*}
    as the right-hand side is a union of countably many $\mathscr{A}\otimes\mathscr{B}_{T}$-measurable sets. Hence the graph $\Gamma(\Phi_{U})$ of $\Phi_{U}$ is $\mathscr{A}\otimes\mathscr{B}_{T}$-measurable. Lemma \ref{lem:relation-bw-two-measurability-notions} shows that $\Phi_{U}$ is also $\mathscr{A}$-measurable.
\end{proof}
\subsection{Measurability Notions for Correspondences: Concrete Results}
We now apply the abstract results of \S\ref{subsec:abstract-corr} to the particular case of correspondences valued in $L^{\infty}(\mathbb{P})$. We equip $L^{\infty}(\mathbb{P})$ with the topology of convergence in probability, and denote by $\mathscr{B}_{p}$ the corresponding Borel $\sigma$-algebra. This may seem like an odd choice, but is justified by a number of facts.
\begin{proposition}\label{prop:number-of-facts-measurability}
    The following holds:
    \begin{enumerate}
        \item\label{fact:metriz-prob} Convergence in probability is metrizable.
        \item\label{fact:local-polish} On every scalar multiple $\lambda B_{L^{\infty}}$ of the unit ball of $L^{\infty}$, convergence in probability is a Polish topology.
        \item\label{fact:souslin-space} $L^{\infty}(\mathbb{P})$ is a Souslin space under convergence in probability.\footnote{Unfortunately, if $\mathbb{P}$ is non-atomic, $L^{\infty}(\mathbb{P})$ cannot be Polish when equipped with convergence in probability. Indeed, the interior of $n B_{L^{\infty}}$ in $L^{\infty}(\mathbb{P})$ with convergence in probability is empty for each $n\in\mathbb{N}$, so $L^{\infty}(\mathbb{P})$ is meagre in itself. However, a non-empty Polish space cannot be meagre in itself by the Baire category theorem.}
        \item\label{fact:measurability-equiv} An allocation $(X_{a})_{a\in A}$ is $\mathscr{A}$-measurable in the sense of Definition \ref{def:meas-weak-star} iff $a\longmapsto X_{a}$ is $(\mathscr{A},\mathscr{B}_{p})$-measurable.
    \end{enumerate}
\end{proposition}
\begin{proof}
    Point \ref{fact:metriz-prob} and Point \ref{fact:local-polish} are obvious. For Point \ref{fact:souslin-space}, consider the Polish space $P=B_{L^{\infty}}\times\mathbb{R}$, where $B_{L^{\infty}}$ is equipped with convergence in probability, and define $h:P\longrightarrow L^{\infty}(\mathbb{P})$ by $h(\mathcal{X},c)=c\mathcal{X}$. Then $h$ is continuous and surjective, so $L^{\infty}(\mathbb{P})$ with convergence in probability is the continuous image of a Polish space—or, what is the same, a Souslin space.
    \par
    Denote by $\mathscr{C}$ the cylindrical $\sigma$-algebra on $L^{\infty}(\mathbb{P})$, which is generated by $L^{1}(\mathbb{P})\subseteq (L^{\infty}(\mathbb{P}))^{\ast}$ (viewed as a set of linear functionals on $L^{\infty}(\mathbb{P})$). Point \ref{fact:measurability-equiv} is equivalent to showing that $\mathscr{B}_{p}=\mathscr{C}$.
    \par
    $\mathscr{C}\subseteq\mathscr{B}_{p}$: Fix an arbitrary $\mathcal{Y}\in L^{1}(\mathbb{P})$, and define a function $\varphi:L^{\infty}(\mathbb{P})\longrightarrow\mathbb{R}$ by $\varphi(\mathcal{X})=\mathbb{E}^{\mathbb{P}}(\mathcal{X}\mathcal{Y})$. It suffices to show that $\varphi$ is $\mathscr{B}_{p}$-measurable. For $n\in\mathbb{N}$, define $\varphi_{n}:L^{\infty}(\mathbb{P})\longrightarrow\mathbb{R}$ by
    \begin{equation*}
        \varphi_{n}(\mathcal{X})=\varphi\left(\left(\mathcal{X}\vee(-n)\right)\wedge n\right).
    \end{equation*}
    By Vitali's convergence theorem, each $\varphi_{n}$ is continuous for convergence in probability, and in particular is $\mathscr{B}_{p}$-measurable. As $\varphi$ is the pointwise limit of the $\varphi_{n}$'s, it follows that $\varphi$ is $\mathscr{B}_{p}$-measurable, as desired.
    \par
    $\mathscr{C}\supseteq\mathscr{B}_{p}$: Denote by $B_{L^{1}}$ the unit ball of $L^{1}(\mathbb{P})$. Fix a sequence $(\mathcal{Y}^{n})_{n=1}^{\infty}\subseteq B_{L^{1}}$ which is norm dense in $B_{L^{1}}$. Define a metric $d$ on $L^{\infty}(\mathbb{P})$ by
    \begin{equation*}
        d\left(\mathcal{X}_{1},\mathcal{X}_{2}\right)=\sum_{n=1}^{\infty}\frac{1}{2^{n}}\left\vert{\mathbb{E}^{\mathbb{P}}\left((\mathcal{X}_{1}-\mathcal{X}_{2})\mathcal{Y}^{n}\right)}\right\vert\wedge1.
    \end{equation*}
    The topology $\tau_{d}$ generated by $d$ is coarser than the weak-star topology on $L^{\infty}(\mathbb{P})$. Denote by $\mathscr{D}$ the Baire $\sigma$-algebra of $\tau_{d}$. Since $\mathscr{C}$ agrees with the Baire $\sigma$-algebra of the weak-star topology (see \cite[Theorem 2.3]{edgar}), it suffices to show that $\mathscr{D}\supseteq\mathscr{B}_{p}$. Remark that, as $\tau_{d}$ is metrizable, $\mathscr{D}$ is also the Borel $\sigma$-algebra $\sigma(\tau_{d})$ of $\tau_{d}$.
    \par
    Fix an arbitrary $U\in\mathscr{B}_{p}$; we must show $U\in\mathscr{D}$. For any $n\in\mathbb{N}$, set $U_{n}=U\cap nB_{L^{\infty}}$, let $\tau_{n,d}$ be the topology generated by $d$ on $nB_{L^{\infty}}$, and let $\tau_{n,p}$ be the topology of convergence in probability on $nB_{L^{\infty}}$. Since $\tau_{n,d}\subseteq\tau_{n,p}$ by Vitali's convergence theorem and both topologies are Polish, $\sigma(\tau_{n,d})=\sigma(\tau_{n,p})$ (see \cite[Exercise 15.4(i)]{kechris}). Since $U_{n}\in\sigma(\tau_{n,p})$, we therefore have that $U_{n}\in\sigma(\tau_{n,d})$. But $\sigma(\tau_{n,d})\subseteq\sigma(\tau_{d})=\mathscr{D}$, so $U_{n}\in\mathscr{D}$. Since $n$ was arbitrary, $U=\bigcup_{n=1}^{\infty}U_{n}\in\mathscr{D}$, as desired.
\end{proof}
Proposition \ref{prop:number-of-facts-measurability} allows the application of the results of \S\ref{subsec:abstract-corr} to $L^{\infty}(\mathbb{P})$.
\begin{proposition}\label{prop:meas-select-specific}
    Let $F:A\longrightarrow 2^{L^{\infty}(\mathbb{P})}$ have an $\mathscr{A}\otimes\mathscr{B}_{p}$-measurable graph. Denote by $B$ the set
    \begin{equation*}
        B=\{a\in A:F(a)\neq\emptyset\}.
    \end{equation*}
    We have that $B\in\mathscr{A}$, and there exists an $\mathscr{A}$-measurable, in the sense of Definition \ref{def:meas-weak-star}, allocation $(X_{a})_{a\in A}$ such that $X_{a}\in F(a)$ for all $a\in B$.
\end{proposition}
\begin{proof}
    Lemma \ref{lem:relation-bw-two-measurability-notions} shows $F$ is $\mathscr{A}$-measurable. By $\mathscr{A}$-measurability of $F$, $B\in\mathscr{A}$. Define a correspondence $G:A\longrightarrow 2^{L^{\infty}(\mathbb{P})}$ by
    \begin{equation*}
        \forall a\in B:G(a)=F(a),
    \end{equation*}
    \begin{equation*}
        \forall a\in A\setminus B:G(a)=L^{\infty}(\mathbb{P}).
    \end{equation*}
    Then
    \begin{equation*}
        \Gamma(G)=\Gamma(F)\cup\left((A\setminus B)\times L^{\infty}(\mathbb{P})\right)\in\mathscr{A}\otimes\mathscr{B}_{p},
    \end{equation*}
    so that $G$ has a measurable graph. By definition, $G(a)\neq\emptyset$ for all $a\in A$. Hence Proposition \ref{prop:measurable-selection-souslin} implies there exists a $(\mathscr{A},\mathscr{B}_{p})$-measurable selector $(X_{a})_{a\in A}$ of $G$. By Fact \ref{fact:measurability-equiv}, $(X_{a})_{a\in A}$ is $\mathscr{A}$-measurable in the sense of Definition \ref{def:meas-weak-star}. If $a\in B$, then $G(a)=F(a)$, so $X_{a}\in F(a)$.
\end{proof}
\begin{proposition}\label{prop:sigma-cara-specific-space}
    Suppose $(V,\mathscr{V})$ is a Polish space with Borel $\sigma$-algebra $\mathscr{B}_{V}$, and let $\Phi:A\times L^{\infty}(\mathbb{P})\longrightarrow V$ be a function with the following properties.
    \begin{enumerate}
        \item For each $a\in A$ and $L^{\infty}(\mathbb{P})$-bounded sequence $(\mathcal{X}^{n})_{n=1}^{\infty}$ converging in probability to some $\mathcal{X}$, we have that
        \begin{equation*}
            \Phi(a,\mathcal{X})=\lim_{n\to\infty}\Phi(a,\mathcal{X}^{n}).
        \end{equation*}
        \item For each $\mathcal{X}\in L^{\infty}(\mathbb{P})$, the map
        \begin{equation*}
            a\longmapsto\Phi(a,\mathcal{X})
        \end{equation*}
        is $(\mathscr{A},\mathscr{B}_{V})$-measurable.
    \end{enumerate}
    For $U\subseteq V$, define $\Phi_{U}:A\longrightarrow 2^{L^{\infty}(\mathbb{P})}$ by
    \begin{equation*}
        \forall a\in A:\Phi_{U}(a)=\left\{\mathcal{X}\in L^{\infty}(\mathbb{P}):\Phi(a,\mathcal{X})\in U\right\}.
    \end{equation*}
    If $U$ is a Borel set in $V$, then the following holds. The correspondence $\Phi_{U}$ has a measurable graph and is $\mathscr{A}$-measurable.
\end{proposition}
\begin{proof}
    The proof follows from verifying the preconditions to Proposition \ref{prop:sigma-cara-corr}. By Point \ref{fact:metriz-prob} and Point \ref{fact:souslin-space} from Proposition \ref{prop:number-of-facts-measurability}, $L^{\infty}(\mathbb{P})$ is a metrizable Souslin space, so all that is left to show is that $\Phi$ is a $\sigma$-Carathéodory function. But this is obvious, since $\Phi|_{A\times T_{n}}$ is a Carathéodory function for each $n\in\mathbb{N}$, where $T_{n}=nB_{L^{\infty}}$.
\end{proof}
\subsection{Measurability Notions for Correspondences: Examples}
We now consider the measurability properties of the specific correspondences we use later on. To this end, let $(\varrho_{a})_{a\in A}$ be an $\mathscr{A}$-measurable set of preferences such that each $\varrho_{a}$ has the Lebesgue property.
\par
Fix a Borel subset $U\subseteq\mathbb{R}$ and an $\mathscr{A}$-measurable function $\xi:A\longrightarrow\mathbb{R}$. Consider the correspondence $F:A\longrightarrow 2^{L^{\infty}(\mathbb{P})}$ defined by
\begin{equation}\label{eq:f-form}
    \forall a\in A:F(a)=\left\{\mathcal{X}\in L^{\infty}(\mathbb{P}):\varrho_{a}(\mathcal{X})-\xi(a)\in U\right\}.
\end{equation}
For example, if $U=(-\infty,0]$ and $\xi=0$, then $F$ is the acceptance set correspondence $a\longmapsto\mathfrak{A}(\varrho_{a})$.
\par
We show that correspondences of the form (\ref{eq:f-form}) satisfy measurability properties. More precisely, we have the following.
\begin{lemma}\label{lem:g-meas}
The correspondence $F$ is $\mathscr{A}$-measurable and has an $\mathscr{A}\otimes\mathscr{B}_{p}$-measurable graph.
\end{lemma}
\begin{proof}
    Define $\Phi:A\times L^{\infty}(\mathbb{P})$ by $\Phi(a,\mathcal{X})=\varrho_{a}(\mathcal{X})-\xi(a)$. Then, by the Lebesgue property of each $\varrho_{a}$ and $\mathscr{A}$-measurability of $(\varrho_{a})_{a\in A}$, $\Phi$ satisfies all preconditions for applying Proposition \ref{prop:sigma-cara-specific-space}. Hence, as $U$ is Borel, the correspondence
    \begin{equation*}
        \Phi_{U}=\left(a\longmapsto\left\{\mathcal{X}\in L^{\infty}(\mathbb{P}):\Phi(a,\mathcal{X})\in U\right\}\right)
    \end{equation*}
    is $\mathscr{A}$-measurable and has an $\mathscr{A}\otimes\mathscr{B}_{p}$-measurable graph. But this correspondence is exactly $F$, showing that $F$ has these properties, as desired.
\end{proof}
Let $G:A\longrightarrow 2^{L^{\infty}(\mathbb{P})}$ be an arbitrary correspondence (although in actual applications $G$ will always take the form (\ref{eq:f-form})). We will later consider ``slices'' of $G$, that is, intersections of $G$ with half-planes.
\begin{definition}
    Given $\mathcal{Y}\in L^{1}(\mathbb{P})$, an $\mathscr{A}$-measurable function $\xi:A\longrightarrow\mathbb{R}$, and a correspondence $G:A\longrightarrow 2^{L^{\infty}(\mathbb{P})}$, the slice correspondence $\mathscr{H}_{G}(\mathcal{Y},\xi)$ is defined by
    \begin{equation*}
        \forall a\in A:\mathscr{H}_{G}(\mathcal{Y},\xi)(a)=\left\{\mathcal{X}\in L^{\infty}(\mathbb{P}):\mathbb{E}^{\mathbb{P}}(\mathcal{X}\mathcal{Y})\geq\xi(a)\right\}\cap G(a).
    \end{equation*}
\end{definition}
\begin{lemma}\label{lem:slice-meas}
    Let $G$ be a correspondence with a measurable graph, and fix some $\mathcal{Y}\in L^{1}(\mathbb{P})$ and some $\mathscr{A}$-measurable $\xi$. Then the slice correspondence $\mathscr{H}_{G}(\mathcal{Y},\xi)$ has a measurable graph and is $\mathscr{A}$-measurable.
\end{lemma}
\begin{proof}
    Let $U=[0,\infty)$. Define a correspondence $H:A\longrightarrow 2^{L^{\infty}(\mathbb{P})}$ by
    \begin{equation*}
        \forall a\in A:H(a)=\left\{\mathcal{X}\in L^{\infty}(\mathbb{P}):\mathbb{E}^{\mathbb{P}}(\mathcal{X}\mathcal{Y})\geq\xi(a)\right\}.
    \end{equation*}
    Since the function $\Phi:A\times L^{\infty}(\mathbb{P})\longrightarrow\mathbb{R}$ defined by $\Phi(a,\mathcal{X})=\mathbb{E}^{\mathbb{P}}(\mathcal{X}\mathcal{Y})-\xi(a)$ satisfies the preconditions of Proposition \ref{prop:sigma-cara-specific-space}, the correspondence $\Phi_{U}:A\longrightarrow L^{\infty}(\mathbb{P})$ defined by
    \begin{equation*}
        \Phi_{U}(a)=\left\{\mathcal{X}\in L^{\infty}(\mathbb{P}):\Phi(a,\mathcal{X})\in U\right\}
    \end{equation*}
    is $\mathscr{A}$-measurable and has a measurable graph. But this correspondence is exactly $H$, so $H$ has these properties.
    \par
    An easy calculation reveals that
    \begin{equation*}
        \Gamma\left(\mathscr{H}_{G}(\mathcal{Y},\xi)\right)=\Gamma(G)\cap\Gamma(H)
    \end{equation*}
    which is evidently $\mathscr{A}\otimes\mathscr{B}_{p}$-measurable. Hence $\mathscr{H}_{G}(\mathcal{Y},\xi)$ has a measurable graph. This also implies $\mathscr{H}_{G}(\mathcal{Y},\xi)$ is $\mathscr{A}$-measurable (see Lemma \ref{lem:relation-bw-two-measurability-notions}).
\end{proof}
\section{Proof of Theorem \ref{thm:finite-conv}}\label{sec:prf-finite-conv}
\subsection{The Richter Theorem}
A classical result of Richter \cite{richter} is that $\mathbb{R}^{d}$-valued Aumann integrals over a non-atomic measure space are convex, which can be viewed as a consequence of Lyapunov convexity.
\begin{lemma}\label{lem:richter}
    Let $F:A\longrightarrow 2^{\mathbb{R}^{d}}$ be a correspondence. The Aumann integral $\int_{A}F(a)\mu(da)$ is convex.
\end{lemma}
Since $\mathscr{F}$ is finite, all the relevant spaces of random variables are finite dimensional, and can therefore be identified with $\mathbb{R}^{d}$ for some dimension $d\in\mathbb{N}$. Identifying $L^{\infty}(\mathbb{P})$ with $\mathbb{R}^{d}$ for some $d\in\mathbb{N}$ allows one to apply Lemma \ref{lem:richter} to $L^{\infty}(\mathbb{P})$-valued correspondences when $\mathscr{F}$ is finite.
\subsection{Proof of Theorem \ref{thm:finite-conv}}
\begin{proof}[Proof of Theorem \ref{thm:finite-conv}]
    Since the integral infimal convolution, when globally finite, is easily seen to be a risk measure, it suffices to show convexity. By \cite[Proposition 4.5(c)]{stoch-fin}, the convexity of a risk measure is equivalent to that of its acceptance set. Theorem \ref{thm:char-acceptance} implies
    \begin{equation*}
        \mathfrak{A}\left(\operatorname{\Box}_{a\in A}\varrho_{a}\mu(da)\right)=\overline{\int_{A}\mathfrak{A}(\varrho_{a})\mu(da)}^{L^{\infty}}
    \end{equation*}
    Since the $L^{\infty}$-closure of a convex set is convex, it suffices to show that $\int_{A}\mathfrak{A}(\varrho_{a})\mu(da)$ is convex, which is a trivial consequence of Lemma \ref{lem:richter}.
\end{proof}
\section{Proof of Theorem \ref{thm:general-conv}}\label{sec:prf-gen-conv}
Before we prove Theorem \ref{thm:general-conv}, let us sketch our general technique. The goal is to apply finite-dimensional Lyapunov convexity by taking conditional expectations with respect to finite sub-$\sigma$-algebras, projecting the relevant random variables onto the finite-dimensional setting of Lemma \ref{lem:richter}, in which everything is convex. After showing these finite-dimensional approximations converge to the proper limit, the claim follows, since convexity is preserved by pointwise limits.
\par
The most clear alternative to the above argument is the use of infinite-dimensional versions of Lyapunov convexity—results similar to Lemma \ref{lem:richter}, but stated for infinite-dimensional spaces. In infinite dimensions, such results generally do not yield convexity of Aumann integrals (see \cite{pod-sat}); rather, they show convexity of weak-star closures of Aumann integrals (see \cite{conv-rep-aumm,kluv,khan85}), and one must therefore assume $\sigma(L^{\infty},L^{1})$-closedness of the acceptance set if they are to be applied in our context.
\par
Unfortunately, a non-convex risk measure $\varrho$ satisfying continuity properties may fail to have a $\sigma(L^{\infty},L^{1})$-closed acceptance set. Indeed, Example \ref{ex:var-counter-integral} below constructs a non-convex risk measure $\varrho$ with the Lebesgue property such that
\begin{equation*}
    \overline{\mathfrak{A}(\varrho)}^{\sigma(L^{\infty},L^{1})}=L^{\infty}(\mathbb{P}).
\end{equation*}
In particular, $\varrho$ is not convex, but $\overline{\mathfrak{A}(\varrho)}^{\sigma(L^{\infty},L^{1})}$ is convex. This implies that if one applied infinite-dimensional Lyapunov convexity theorems, the resulting conclusion—that the weak-star closure of the acceptance set is convex—does not in general imply the acceptance set is convex. In other optimization contexts, similar concerns imply aggregate convexity applies not to the original value function, but to some closed hull thereof (see \cite[\S5]{conv-rep-aumm}). In particular, existing approaches based on infinite-dimensional Lyapunov convexity are insufficient for our purposes.
\begin{example}\label{ex:var-counter-integral}
    Assume $(\Omega,\mathscr{F},\mathbb{P})$ is non-atomic. Define the risk measure $\varrho$ by
    \begin{equation*}
        \forall\mathcal{X}\in L^{\infty}(\mathbb{P}):\varrho(\mathcal{X})=2\int_{1/2}^{1}\mathrm{VaR}_{\beta}(\mathcal{X})d\beta,
    \end{equation*}
    which has the Lebesgue property and is law invariant, where
    \begin{equation*}
        \forall\mathcal{X}\in L^{\infty}(\mathbb{P}):\textrm{VaR}_{\beta}(\mathcal{X})=\inf\left\{x:\mathbb{P}(\{X\leq x\})\geq1-\beta\right\}
    \end{equation*}
    is the Value-at-Risk at level $\beta\in(0,1)$. Defining
    \begin{equation*}
        \mathcal{A}= \overline{\mathfrak{A}(\varrho)}^{\sigma(L^{\infty},L^{1})}
    \end{equation*}
    we claim that $\mathcal{A}=L^{\infty}(\mathbb{P})$.
    \par
    Let $\mathcal{X}\in L^{\infty}(\mathbb{P})$. Find $\mathcal{X}'\sim\mathcal{X}$ such that there exists $\mathcal{X}''$ independent of $\mathcal{X}'$ with $\mathcal{X}''\sim\frac{3}{4}\delta_{0}+\frac{1}{4}\delta_{4}$. Put $\mathcal{Y}=\mathcal{X}'\mathcal{X}''$. Since $\mathbb{P}\left(\left\{\mathcal{Y}\leq0\right\}\right)\geq\frac{3}{4}$, it follows from an easy calculation that $\varrho\left(\mathcal{Y}\right)\leq0$. As is easily seen, $\mathcal{Y}$ dominates $\mathcal{X}$ in the convex order. Hence, by Ryff's theorem \cite[Proposition 15]{sch-ivar}, there exists a sequence $({\mathcal{Y}}^{n})_{n=1}^{\infty}$ converging in $\sigma(L^{1},L^{\infty})$ to $\mathcal{X}$ such that $\mathcal{Y}^{n}\sim\mathcal{Y}$ for each $n\in\mathbb{N}$. Since the sequence $(\mathcal{Y}^{n})_{n=1}^{\infty}$ is $L^{\infty}(\mathbb{P})$ bounded, convergence to $\mathcal{X}$ occurs in $\sigma(L^{\infty},L^{1})$. As $\varrho$ is law invariant and $\varrho\left(\mathcal{Y}\right)\leq0$, we must have $\varrho\left(\mathcal{Y}^{n}\right)\leq0$ for each $n$, so that $\mathcal{Y}^{n}\in\mathfrak{A}(\varrho)$ for each $n$. Hence $\mathcal{X}$ lies in the weak-star closure $\mathcal{A}$ of $\mathfrak{A}(\varrho)$. Since $\mathcal{X}\in L^{\infty}(\mathbb{P})$ was arbitrary, it follows that $\mathcal{A}=L^{\infty}(\mathbb{P})$, as desired.
    \qed
\end{example}
\subsection{The Infimal $\mathscr{G}$-Convolution}\label{subsec:g-convolve}
Let $\mathscr{G}\subseteq\mathscr{F}$ be a sub-$\sigma$-algebra. Denote by $L^{\infty}(\mathscr{G},\mathbb{P})$ the subspace of $L^{\infty}(\mathbb{P})$ consisting of elements $\mathcal{X}\in L^{\infty}(\mathbb{P})$ with a $\mathscr{G}$-measurable $\mathbb{P}$-modification. It is not difficult to see that $L^{\infty}(\mathscr{G},\mathbb{P})$ is canonically identifiable with the $L^{\infty}$-space associated to the probability space $(\Omega,\mathscr{G},\mathbb{P}|_{\mathscr{G}})$, and we therefore do not distinguish between these two spaces.
\begin{definition}
    Suppose $\mathcal{X}\in L^{\infty}(\mathscr{G},\mathbb{P})$. An allocation $(X_{a})_{a\in A}\in\mathbb{A}(\mathcal{X})$ is $\mathscr{G}$-feasible if $X_{a}\in L^{\infty}(\mathscr{G},\mathbb{P})$ for $\mu$-a.e. $a\in A$. The set of $\mathscr{G}$-feasible allocations of $\mathcal{X}$ is denoted $\mathbb{A}(\mathscr{G},\mathcal{X})$.
\end{definition}
\begin{definition}
    The integral infimal $\mathscr{G}$-convolution $\mathscr{G}\textrm{-}\operatorname{\Box}_{a\in A}\varrho_{a}\mu(da)$ is defined as
    \begin{equation*}
        \left(\mathscr{G}\textrm{-}\operatorname{\Box}_{a\in A}\varrho_{a}\mu(da)\right)(\mathcal{X})=\inf_{(X_{a})_{a\in A}\in\mathbb{A}(\mathscr{G},\mathcal{X})}\int_{A}\varrho_{a}(X_{a})\mu(da).
    \end{equation*}
    for each $\mathcal{X}\in L^{\infty}(\mathscr{G},\mathbb{P})$.
\end{definition}
\begin{remark}\label{rem:apply-finite-conv}
    Suppose $\mathscr{G}$ is a finite sub-$\sigma$-algebra of $\mathscr{F}$. If $\mathscr{G}\textrm{-}\operatorname{\Box}_{a\in A}\varrho_{a}\mu(da)$ is globally finite, Theorem \ref{thm:finite-conv} applies to show that $\mathscr{G}\textrm{-}\operatorname{\Box}_{a\in A}\varrho_{a}\mu(da)$ is convex.
    \qed
\end{remark}
\begin{lemma}\label{lem:gconvolve-extension}
    Let $\mathbb{Q}\sim\mathbb{P}$. Suppose that, for each $a\in A$, $\varrho_{a}$ is $\mathbb{Q}$-consistent. Then
    \begin{equation*}
        \mathscr{G}\textrm{-}\operatorname{\Box}_{a\in A}\varrho_{a}\mu(da)=\operatorname{\Box}_{a\in A}\varrho_{a}\mu(da)|_{L^{\infty}(\mathscr{G},\mathbb{P})}.
    \end{equation*}
\end{lemma}
\begin{proof}
    Let $\mathcal{X}\in L^{\infty}(\mathscr{G},\mathbb{P})$ be arbitrary. For any allocation $(X_{a})_{a\in A}\in\mathbb{A}(\mathcal{X})$, $(Y_{a})_{a\in A}$ defined by $Y_{a}=\mathbb{E}^{\mathbb{Q}}(X_{a}|\mathscr{G})$ is easily seen to be a $\mathscr{G}$-feasible allocation of $\mathcal{X}$. By Proposition \ref{prop:dm-iff-cons}, each $\varrho_{a}$ is $\mathbb{Q}$-dilatation monotone. Thus,
    \begin{equation*}
        \varrho_{a}(Y_{a})\leq\varrho_{a}(X_{a})
    \end{equation*}
    for each $a\in A$, implying $\mathscr{G}\textrm{-}\operatorname{\Box}_{a\in A}\varrho_{a}\mu(da)\leq\operatorname{\Box}_{a\in A}\varrho_{a}\mu(da)|_{L^{\infty}(\mathscr{G},\mathbb{P})}$. The reverse inequality is a trivial consequence of the inclusion $\mathbb{A}(\mathscr{G},\mathcal{X})\subseteq\mathbb{A}(\mathcal{X})$.
\end{proof}
\subsection{Stability of the Value Function}\label{subsec:stability-val}
In this subsection, we consider to what extent the integral infimal convolution inherits properties from $(\varrho_{a})_{a\in A}$. Throughout this subsection, we assume that each $\varrho_{a}$ has the Lebesgue property and is $\mathbb{Q}$-consistent for some fixed $\mathbb{Q}\sim\mathbb{P}$ (this corresponds to the case $\pi=\{A\}$).
\par
The following is a generalization of a known result for the ordinary infimal convolution (see \cite[Example 3]{law-inv-non-conv}):
\begin{lemma}\label{lem:consistent-stable}
    The integral infimal convolution $\operatorname{\Box}_{a\in A}\varrho_{a}\mu(da)$ is $\mathbb{Q}$-consistent.
\end{lemma}
\begin{proof}
    The argument is essentially the same as Lemma \ref{lem:gconvolve-extension}.
\end{proof}
\begin{lemma}\label{lem:lebesgue-stable}
    The integral infimal convolution $\operatorname{\Box}_{a\in A}\varrho_{a}\mu(da)$ satisfies the Lebesgue property.
\end{lemma}
\begin{proof}
    The Lebesgue property is equivalent to continuity from above and from below. By Lemma \ref{lem:consistent-stable}, $\operatorname{\Box}_{a\in A}\varrho_{a}\mu(da)$ is $\mathbb{Q}$-consistent, and Proposition \ref{prop:cons-fatou} therefore ensures $\operatorname{\Box}_{a\in A}\varrho_{a}\mu(da)$ has the Fatou property. The Fatou property implies continuity from below, and it therefore suffices to show that $\operatorname{\Box}_{a\in A}\varrho_{a}\mu(da)$ is continuous from above. Continuity from above follows in a standard way from the monotone convergence theorem and the Lebesgue property of each $\varrho_{a}$ (see \cite[Proposition 2]{melnikov-general-theory}).
\end{proof}
\subsection{Partitioned Infimal Convolutions}\label{subsec:partition-prf}
In this subsection, we will use the notation $\varrho_{B}$ for the integral infimal convolution $\operatorname{\Box}_{a\in B}\varrho_{a}\mu(da)$ for $B\in\pi$. The agent space of this integral infimal convolution is $(B,\mathscr{A}_{B},\mu|_{B})$ ($\mathscr{A}_{B}$ is the trace $\sigma$-algebra of $\mathscr{A}$ on $B$). In particular, denoting by $\mathbb{A}_{B}(\mathcal{X})$ the space of allocations of $\mathcal{X}$ on the measure space $(B,\mathscr{A}_{B},\mu|_{B})$, we have that
\begin{equation*}
    \varrho_{B}(\mathcal{X})=\inf_{(X_{a})_{a\in B}\in\mathbb{A}_{B}(\mathcal{X})}\int_{B}\varrho_{a}(X_{a})\mu(da)
\end{equation*}
for each $B\in\pi$ and $\mathcal{X}\in L^{\infty}(\mathbb{P})$. Similarly, we use $\mathbb{A}_{\pi}(\mathcal{X})$ for the space of allocations of $\mathcal{X}$ on the agent space
\begin{equation*}
    \left(\pi,2^{\pi},\sum_{B\in\pi}\delta_{B}\right)
\end{equation*}
where $\delta_{B}$ denotes the Dirac measure centered at $B\in\pi$, and denote by $\varrho_{\pi}=\operatorname{\Box}_{B\in\pi}\varrho_{B}$ the corresponding infimal convolution.
\begin{lemma}\label{lem:isriskmeasure}
    For each $B\in\pi$, $\varrho_{B}$ is a risk measure, and in particular is globally finite.
\end{lemma}
\begin{proof}
    Monotonicity and cash additivity follow easily as long as the relevant functionals are globally finite, and so we focus on global finiteness. We have that $\varrho_{B}(\mathcal{X})\leq\Vert{\mathcal{X}}\Vert_{L^{\infty}}+\int_{B}\vert{\varrho_{a}}(0)\vert\mu(da)<\infty$ for any $\mathcal{X}$. So it suffices to show that $\varrho_{B}$ is not identically $-\infty$ (see Proposition \ref{prop:basic-properties}).
    \par
    If $\varrho_{B}(0)=-\infty$, there exists a sequence $\left((X^{n}_{a})_{a\in B}\right)_{n=1}^{\infty}\subseteq\mathbb{A}_{B}(0)$ such that
    \begin{equation}\label{eq:neg-inf-partition}
        \inf_{n}\int_{B}\varrho_{a}(X^{n}_{a})\mu(da)=-\infty.
    \end{equation}
    For each $n\in\mathbb{N}$, define $(Y^{n}_{a})_{a\in A}\in\mathbb{A}(0)$ by setting $Y^{n}_{a}=X^{n}_{a}$ for $a\in B$, and $Y^{n}_{a}=0$ for $a\in A\setminus B$. Since $\int_{A\setminus B}\vert{\varrho_{a}(0)}\vert\mu(da)<\infty$, (\ref{eq:neg-inf-partition}) implies
    \begin{equation*}
        \left(\operatorname{\Box}_{a\in A}\varrho_{a}\mu(da)\right)(0)\leq\inf_{n}\int_{A}\varrho_{a}(Y^{n}_{a})\mu(da)=-\infty.
    \end{equation*}
    In particular, $\left(\operatorname{\Box}_{a\in A}\varrho_{a}\mu(da)\right)(0)=-\infty$, contradicting global finiteness of the integral infimal convolution.
\end{proof}
\begin{lemma}\label{lem:partition-convolve}
    We have that
    \begin{equation*}
        \operatorname{\Box}_{a\in A}\varrho_{a}\mu(da)=\varrho_{\pi}.
    \end{equation*}
\end{lemma}
Since the infimal convolution of finitely many convex risk measures is convex and inherits the Lebesgue property whenever globally finite, Lemma \ref{lem:partition-convolve} implies that, for the proof of Theorem \ref{thm:general-conv}, we may reduce to the case $\pi=\{A\}$.
\begin{proof}
    Let $\mathcal{X}\in L^{\infty}(\mathbb{P})$ be arbitrary. For each $(X_{a})_{a\in A}\in\mathbb{A}(\mathcal{X})$,
    \begin{equation*}
        \int_{A}\varrho_{a}(X_{a})\mu(da)\geq\sum_{B\in\pi}\varrho_{B}\left(\int_{B}X_{a}\mu(da)\right)\geq\varrho_{\pi}\left(\sum_{B\in\pi}\int_{B}X_{a}\mu(da)\right)=\varrho_{\pi}(\mathcal{X}),
    \end{equation*}
    so that, after taking the infimum over $(X_{a})_{a\in A}\in\mathbb{A}(\mathcal{X})$,
    \begin{equation*}
        \left(\operatorname{\Box}_{a\in A}\varrho_{a}\mu(da)\right)(\mathcal{X})\geq\varrho_{\pi}(\mathcal{X}).
    \end{equation*}
    Thus, it suffices to show that there cannot exist $\delta>0$ with
    \begin{equation}\label{eq:finite-inf-better}
        \left(\operatorname{\Box}_{a\in A}\varrho_{a}\mu(da)\right)(\mathcal{X})-\delta>\varrho_{\pi}(\mathcal{X}).
    \end{equation}
    Suppose (\ref{eq:finite-inf-better}) holds. We may find $(X_{B})_{B\in\pi}\in\mathbb{A}_{\pi}(\mathcal{X})$ with
    \begin{equation*}
        \left(\operatorname{\Box}_{a\in A}\varrho_{a}\mu(da)\right)(\mathcal{X})-\delta>\sum_{B\in\pi}\varrho_{B}(X_{B}).
    \end{equation*}
    By Lemma \ref{lem:isriskmeasure}, for each $\varepsilon>0$ (we later take $\varepsilon<\delta/\vert{\pi}\vert$) and $B\in\pi$, we may find $(X^{B,\varepsilon}_{a})_{a\in B}\in\mathbb{A}_{B}(X_{B})$ such that
    \begin{equation*}
        \int_{B}\varrho_{a}(X^{B,\varepsilon}_{a})\mu(da)\leq\varrho_{B}(X_{B})+\varepsilon.
    \end{equation*}
    Define $(Y^{\varepsilon}_{a})_{a\in A}\in\mathbb{A}(\mathcal{X})$ by setting $Y^{\varepsilon}_{a}=X^{B,\varepsilon}_{a}$ for $a\in B$ for each $B\in\pi$. Then,
    \begin{equation*}
        \sum_{B\in\pi}\varrho_{B}(X_{B})+\delta<\left(\operatorname{\Box}_{a\in A}\varrho_{a}\mu(da)\right)(\mathcal{X})\leq\int_{A}\varrho_{a}(Y^{\varepsilon}_{a})\mu(da)\leq\sum_{B\in\pi}\varrho_{B}(X_{B})+\varepsilon\vert{\pi}\vert.
    \end{equation*}
    Taking $\varepsilon\to0$ so that $\delta-\varepsilon\vert{\pi}\vert>0$ yields a contradiction.
\end{proof}
\subsection{Proof of Theorem \ref{thm:general-conv}}\label{subsec:prf-general-conv}
\begin{proof}[Proof of Theorem \ref{thm:general-conv}]
    By Lemma \ref{lem:isriskmeasure} and Lemma \ref{lem:partition-convolve}, it suffices to prove the claim for the singleton $\pi=\{A\}$ and a single probability measure $\mathbb{Q}\sim\mathbb{P}$. Thus, we may assume, without loss of generality, that each $\varrho_{a}$ is $\mathbb{Q}$-consistent with respect to a fixed $\mathbb{Q}\sim\mathbb{P}$.
    \par
    Since the integral infimal convolution, when globally finite, is a risk measure (see Proposition \ref{prop:basic-properties}), it suffices to show convexity and the Lebesgue property. The Lebesgue property follows from Lemma \ref{lem:lebesgue-stable}, so we focus now on convexity.
    \par
    Let $(\mathscr{G}_{n})_{n=1}^{\infty}$ be an increasing sequence of finite sub-$\sigma$-algebras of $\mathscr{F}$ such that $\bigcup_{n=1}^{\infty}\mathscr{G}_{n}$ is $\mathbb{P}$-dense in $\mathscr{F}$. Write $P_{n}=\mathbb{E}^{\mathbb{Q}}(\cdot|\mathscr{G}_{n})$ for the conditional $\mathbb{Q}$-expectation with respect to $\mathscr{G}_{n}$. Consider the sequence $(\varrho^{n})_{n=1}^{\infty}$ of risk measures, where
    \begin{equation*}
        \varrho^{n}=\left(\mathscr{G}_{n}\textrm{-}\operatorname{\Box}_{a\in A}\varrho_{a}\mu(da)\right)\circ P_{n}
    \end{equation*}
    for each $n\in\mathbb{N}$. By Remark \ref{rem:apply-finite-conv}, $\mathscr{G}_{n}\textrm{-}\operatorname{\Box}_{a\in A}\varrho_{a}\mu(da)$ is convex, implying $\varrho^{n}$ is convex. Lemma \ref{lem:gconvolve-extension} implies that
    \begin{equation}\label{eq:composition-with-expectation}
        \varrho^{n}=\left(\operatorname{\Box}_{a\in A}\varrho_{a}\mu(da)\right)\circ P_{n}
    \end{equation}
    for each $n\in\mathbb{N}$. By Doob's martingale convergence theorem, if $\mathcal{X}\in L^{\infty}(\mathbb{P})$, we have that the sequence $(P_{n}(\mathcal{X}))_{n=1}^{\infty}$ is $L^{\infty}(\mathbb{P})$-bounded and converges in probability to $\mathcal{X}$. From (\ref{eq:composition-with-expectation}) and the Lebesgue property of the integral infimal convolution, we thus get that
    \begin{equation*}
        \forall\mathcal{X}\in L^{\infty}(\mathbb{P}):\left(\operatorname{\Box}_{a\in A}\varrho_{a}\mu(da)\right)(\mathcal{X})=\lim_{n\to\infty}\varrho^{n}(\mathcal{X}).
    \end{equation*}
    Hence the integral infimal convolution is a pointwise limit of convex functions, which proves convexity.
\end{proof}
\section{Proof of Theorem \ref{thm:conv-conj-value}}\label{sec:prf-formula-conj}
\subsection{Representing the Dual via Acceptance Sets}
We first convert the problem of characterizing the convex conjugate of a risk measure into a more tractable problem involving acceptance sets.
\begin{lemma}\label{lem:sup-over-acc-set}
    Let $\varrho$ be a risk measure. Then, for any $\mathbb{Q}\in\mathscr{M}_{\mathbb{P}}$,
    \begin{equation*}
        \varrho^{\ast}(\mathbb{Q})=\sup_{\mathcal{X}\in\mathfrak{A}(\varrho)}\mathbb{E}^{\mathbb{Q}}(\mathcal{X}).
    \end{equation*}
\end{lemma}
\begin{proof}
    This is a well-known fact (see \cite[Lemma 3]{melnikov-general-theory}).
\end{proof}
\subsection{Proof of Theorem \ref{thm:conv-conj-value}}
We deduce Theorem \ref{thm:conv-conj-value} from a more abstract formulation.
\begin{theorem}\label{thm:conj-abstract}
    Let $F:A\longrightarrow 2^{L^{\infty}(\mathbb{P})}$ have non-empty values, an $\mathscr{A}\otimes\mathscr{B}_{p}$-measurable graph, and $S^{1}(F)\neq\emptyset$. Fix $\mathcal{Y}\in L^{1}(\mathbb{P})$. Denoting
    \begin{equation*}
        \Xi_{\mathcal{Y}}=\left(a\longmapsto\sup_{\mathcal{X}\in F(a)}\mathbb{E}^{\mathbb{P}}(\mathcal{X}\mathcal{Y})\right),
    \end{equation*}
    we have that $\Xi_{\mathcal{Y}}$ is $\mathscr{A}$-measurable, and
    \begin{equation}\label{eq:abstract-sup-switch}
        \sup_{\mathcal{X}\in\int_{A}F(a)\mu(da)}\mathbb{E}^{\mathbb{P}}(\mathcal{X}\mathcal{Y})=\int_{A}\Xi_{\mathcal{Y}}(a)\mu(da).
    \end{equation}
\end{theorem}
\begin{remark}
To see that Theorem \ref{thm:conj-abstract} implies Theorem \ref{thm:conv-conj-value}, put $F(a)=\mathfrak{A}(\varrho_{a})$ and $\mathcal{Y}=\frac{d\mathbb{Q}}{d\mathbb{P}}$. Then $F$ satisfies all the preconditions for Theorem \ref{thm:conj-abstract} (indeed, $(-\varrho_{a}(0))_{a\in A}\in S^{1}(F)$, and Proposition \ref{prop:sigma-cara-specific-space} shows $F$ has a measurable graph), and so applying it yields
\begin{equation}\label{eq:first-equality-conj}
    \sup_{\mathcal{X}\in\int_{A}\mathfrak{A}(\varrho_{a})\mu(da)}\mathbb{E}^{\mathbb{Q}}(\mathcal{X})=\int_{A}\sup_{\mathcal{X}\in\mathfrak{A}(\varrho_{a})}\mathbb{E}^{\mathbb{Q}}(\mathcal{X})\mu(da)=\int_{A}\varrho^{\ast}_{a}(\mathbb{Q})\mu(da),
\end{equation}
where the second equality follows from Lemma \ref{lem:sup-over-acc-set}. But Lemma \ref{lem:sup-over-acc-set} together with Theorem \ref{thm:char-acceptance} implies
\begin{equation}\label{eq:second-equality-conj}
    \left(\operatorname{\Box}_{a\in A}\varrho_{a}\mu(da)\right)^{\ast}(\mathbb{Q})=\sup_{\mathcal{X}\in\overline{\int_{A}\mathfrak{A}(\varrho_{a})\mu(da)}}\mathbb{E}^{\mathbb{Q}}(\mathcal{X})=\sup_{\mathcal{X}\in \int_{A}\mathfrak{A}(\varrho_{a})\mu(da)}\mathbb{E}^{\mathbb{Q}}(\mathcal{X}),
\end{equation}
where the closure is taken in $L^{\infty}$-norm, and the second equality follows from norm continuity of the map $\mathcal{X}\longmapsto\mathbb{E}^{\mathbb{Q}}(\mathcal{X})$. Combining (\ref{eq:first-equality-conj}) and (\ref{eq:second-equality-conj}) yields
\begin{equation*}
    \left(\operatorname{\Box}_{a\in A}\varrho_{a}\mu(da)\right)^{\ast}(\mathbb{Q})=\int_{A}\varrho^{\ast}_{a}(\mathbb{Q})\mu(da),
\end{equation*}
as desired.
\qed
\end{remark}
The first step in our proof of Theorem \ref{thm:conj-abstract} is deducing $\mathscr{A}$-measurability of $\Xi_{\mathcal{Y}}$.
\begin{lemma}\label{lem:sup-is-measurable}
    $\Xi_{\mathcal{Y}}$ is $\mathscr{A}$-measurable.
\end{lemma}
\begin{proof}
    We equip $L^{\infty}(\mathbb{P})$ with the Mackey topology $\mathfrak{m}=\tau(L^{\infty},L^{1})$; for a full definition of the Mackey topology, see \cite[\S5.18]{aliprantis-inf-dim}. We list here only some relevant properties of $\mathfrak{m}$.
\begin{enumerate}
    \item\label{mackey:cont} The map $\mathcal{X}\longmapsto\mathbb{E}^{\mathbb{P}}(\mathcal{X}\mathcal{Y})$ is $\mathfrak{m}$-continuous (see \cite[\S5.18]{aliprantis-inf-dim}).
    \item\label{mackey:finer} The topology $\mathfrak{m}$ is finer than the topology of convergence in probability (see \cite[pp. 216-223]{gro-tvs}). In particular, the unit ball $B_{L^{\infty}}$, and every scalar multiple thereof, is $\mathfrak{m}$-closed.
    \item\label{mackey:coincide} For every $L^{\infty}$-bounded $U\subseteq L^{\infty}(\mathbb{P})$, the subspace topology for $\mathfrak{m}$ agrees with the subspace topology for convergence in probability (see \cite[pp. 216-223]{gro-tvs}).
    \item\label{mackey:souslin} A consequence of the previous two points is that $(L^{\infty},\mathfrak{m})$ is a Souslin space (use the same argument as given for Point \ref{fact:souslin-space} of Proposition \ref{prop:number-of-facts-measurability}), and the Borel $\sigma$-algebra $\sigma(\mathfrak{m})$ coincides with $\mathscr{B}_{p}$ (indeed, in view of Point \ref{mackey:finer} and Point \ref{mackey:coincide} directly above, $U\in\sigma(\mathfrak{m})\iff\forall n\in\mathbb{N}:U\cap nB_{L^{\infty}}\in\sigma(\mathfrak{m})\iff\forall n\in\mathbb{N}:U\cap nB_{L^{\infty}}\in\mathscr{B}_{p}\iff U\in\mathscr{B}_{p}$).
\end{enumerate}
The last point implies $F$ has an $\mathscr{A}\otimes\sigma(\mathfrak{m})$-measurable graph, and Proposition \ref{prop:measurable-selection-souslin} thus implies the existence of a sequence  $\left((X^{n}_{a})_{a\in A}\right)_{n=1}^{\infty}$ of $(\mathscr{A},\sigma(\mathfrak{m}))$-measurable selectors of $F$ with $\{X^{n}_{a}:n\in\mathbb{N}\}\subseteq F(a)$ $\mathfrak{m}$-dense in $F(a)$ for each $a\in A$. Because of $\mathfrak{m}$-continuity of $\mathcal{X}\longmapsto\mathbb{E}^{\mathbb{P}}(\mathcal{X}\mathcal{Y})$,
\begin{equation*}
    \Xi_{\mathcal{Y}}(a)=\sup_{\mathcal{X}\in F(a)}\mathbb{E}^{\mathbb{P}}(\mathcal{X}\mathcal{Y})=\sup_{n\in\mathbb{N}}\mathbb{E}^{\mathbb{P}}(X^{n}_{a}\mathcal{Y})
\end{equation*}
for each $a\in A$, representing $\Xi_{\mathcal{Y}}$ as a countable supremum of $\mathscr{A}$-measurable functions.
\end{proof}
We are now ready to prove Theorem \ref{thm:conj-abstract}.
\begin{proof}[Proof of Theorem \ref{thm:conj-abstract}]
    The $\mathscr{A}$-measurability of $\Xi_{\mathcal{Y}}$ follows from Lemma \ref{lem:sup-is-measurable}. Suppose, for contradiction, that (\ref{eq:abstract-sup-switch}) fails, in which case there exists a $\mu$-integrable function $\xi\leq\Xi_{\mathcal{Y}}$ and $\delta>0$ with
    \begin{equation}\label{eq:failure-of-sup-switch}
        \sup_{\mathcal{X}\in\int_{A}F(a)\mu(da)}\mathbb{E}^{\mathbb{P}}(\mathcal{X}\mathcal{Y})<\int_{A}\xi(a)\mu(da)-\delta.
    \end{equation}
    Define $\mathfrak{F}:A\longrightarrow 2^{L^{\infty}(\mathbb{P})}$ by
    \begin{equation*}
        \forall a\in A:\mathfrak{F}(a)=F(a)\cap\left\{\mathcal{X}:\mathbb{E}^{\mathbb{P}}(\mathcal{X})\geq\xi(a)-\frac{\delta}{4\mu(A)}\right\}
    \end{equation*}
    As $\xi\leq\Xi_{\mathcal{Y}}$, $\mathfrak{F}(a)\neq\emptyset$ for all $a$. Remark that $\mathfrak{F}$ is a slice correspondence: $\mathfrak{F}=\mathscr{H}_{F}(\mathcal{Y},\zeta)$ for $\zeta=\xi-\frac{\delta}{4\mu(A)}$. Hence $\mathfrak{F}$ inherits the property of having a measurable graph from $F$ (see Lemma \ref{lem:slice-meas}). In particular, Proposition \ref{prop:meas-select-specific} implies that there exists an $\mathscr{A}$-measurable selector $(Z_{a})_{a\in A}$ of $\mathfrak{F}$.
    \par
    Fix $W\in S^{1}(F)$. For $n\in\mathbb{N}$, define $B^{n}=\left\{a\in A:\left\Vert{Z_{a}}\right\Vert_{L^{\infty}}\leq n\right\}$ and
    \begin{equation*}
        Z^{n}=\mathbf{1}_{B^{n}}Z+\mathbf{1}_{A\setminus B^{n}}W.
    \end{equation*}
    Then $Z^{n}\in S^{1}(F)$. We have that
    \begin{equation*}
        \int_{A}\mathbb{E}^{\mathbb{P}}(Z^{n}_{a}\mathcal{Y})\mu(da)\geq\int_{B^{n}}\left(\xi(a)-\frac{\delta}{4\mu(A)}\right)\mu(da)+\int_{A\setminus B^{n}}\mathbb{E}^{\mathbb{P}}(W_{a}\mathcal{Y})
    \end{equation*}
    \begin{equation*}
        =-\frac{\delta}{4}+\int_{B^{n}}\xi(a)\mu(da)+\int_{A\setminus B^{n}}\mathbb{E}^{\mathbb{P}}(W_{a}\mathcal{Y}).
    \end{equation*}
    For large enough $n$ (say, $n\geq n_{0}$), we have that
    \begin{equation*}
        \int_{B^{n}}\xi(a)\mu(da)\geq-\frac{\delta}{4}+\int_{A}\xi(a)\mu(da)\textrm{ and }\int_{A\setminus B^{n}}\mathbb{E}^{\mathbb{P}}(W_{a}\mathcal{Y})\mu(da)\geq-\frac{\delta}{4},
    \end{equation*}
    as a consequence of the dominated convergence theorem (the $B^{n}$'s form an increasing sequence covering $A$). Hence, in view of (\ref{eq:failure-of-sup-switch}) and the fact that $Z^{n_{0}}\in S^{1}(F)$,
    \begin{equation*}
        \int_{A}\mathbb{E}^{\mathbb{P}}(Z^{n_{0}}_{a}\mathcal{Y})\mu(da)\geq-\frac{\delta}{4}+\int_{B^{n_{0}}}\xi(a)\mu(da)+\int_{A\setminus B^{n_{0}}}\mathbb{E}^{\mathbb{P}}(W_{a}\mathcal{Y})\geq-\frac{3\delta}{4}+\int_{A}\xi(a)\mu(da)
    \end{equation*}
    \begin{equation*}
        \geq\frac{\delta}{4}+\int_{A}\mathbb{E}^{\mathbb{P}}(Z^{n_{0}}_{a}\mathcal{Y})\mu(da),
    \end{equation*}
    a contradiction.
\end{proof}
\section{Proof of Theorem \ref{thm:welfare-theorems}}\label{sec:prf-welfare}
\subsection{Subgradients and the Lebesgue Property}
\begin{definition}
    Let $\varrho:L^{\infty}(\mathbb{P})\longrightarrow\mathbb{R}\cup\{\infty\}$ be a proper functional, and fix some $\mathcal{X}_{0}\in L^{\infty}(\mathbb{P})$ with $\varrho(\mathcal{X}_{0})<\infty$. A subgradient of $\varrho$ at $\mathcal{X}_{0}$ is a point $\mathcal{Y}\in L^{1}(\mathbb{P})$ such that, for all $\mathcal{X}\in L^{\infty}(\mathbb{P})$,
    \begin{equation*}
        \varrho(\mathcal{X})-\varrho(\mathcal{X}_{0})\geq\mathbb{E}^{\mathbb{P}}(\mathcal{Y}(\mathcal{X}-\mathcal{X}_{0})).
    \end{equation*}
    The set of all subgradients of $\varrho$ at $\mathcal{X}_{0}$ is denoted $\partial \varrho(\mathcal{X}_{0})$, and is called the subdifferential of $\varrho$ at $\mathcal{X}_{0}$.
\end{definition}
The following result is a consequence of the Jouini-Schachermayer-Touzi theorem (see \cite[Theorem 5.2]{law-inv-util-conv-dual}).
\begin{proposition}\label{prop:jst-subdifferential}
    Let $\varrho:L^{\infty}(\mathbb{P})\longrightarrow\mathbb{R}$ be a convex risk measure with the Lebesgue property. Then $\partial\varrho(\mathcal{X}_{0})\neq\emptyset$ for all $\mathcal{X}_{0}\in L^{\infty}(\mathbb{P})$, and $\partial\varrho(\mathcal{X}_{0})\subseteq\mathscr{M}_{\mathbb{P}}$.
\end{proposition}
We show in the proof of Theorem \ref{thm:welfare-theorems} that the subdifferential of the integral infimal convolution consists of possible equilibrium pricing vectors (cf. Theorem 5.5, \cite{subgrad_law_inv}). In light of Proposition \ref{prop:jst-subdifferential}, for this argument to work, it is critical that the integral infimal convolution has the Lebesgue property.
\begin{lemma}\label{lem:has-lebesgue-equi}
    The integral infimal convolution $\operatorname{\Box}_{a\in A}\varrho_{a}\mu(da)$ has the Lebesgue property.
\end{lemma}
\begin{proof}
    It is assumed in Theorem \ref{thm:welfare-theorems} that $\operatorname{\Box}_{a\in A}\varrho_{a}\mu(da)$ is convex, and each $\varrho_{a}$ has the Lebesgue property. By \cite[Remark 4.19]{stoch-fin}, the Lebesgue property for $\operatorname{\Box}_{a\in A}\varrho_{a}\mu(da)$ is thus equivalent to continuity from above. Continuity from above follows in a standard way from the monotone convergence theorem and the Lebesgue property of each $\varrho_{a}$ (see \cite[Proposition 2]{melnikov-general-theory}).
\end{proof}
\subsection{Proof of Theorem \ref{thm:welfare-theorems}}\label{subsec:prf-welfare}
\begin{proof}
    Let $(X_{a})_{a\in A}$ be Pareto efficient, and define $(Y_{a})_{a\in A}$ by $Y_{a}=X_{a}+\mathbb{E}^{\mathbb{Q}}(E_{a}-X_{a})$. Remark that the Pareto efficiency of $(X_{a})_{a\in A}$, together with the assumption that $\left(\operatorname{\Box}_{a\in A}\varrho_{a}\mu(da)\right)(\mathcal{X})<\infty$, imply that $\left(\operatorname{\Box}_{a\in A}\varrho_{a}\mu(da)\right)(\mathcal{X})=\int_{A}\varrho_{a}(X_{a})\mu(da)$ (see Proposition \ref{prop:pareto-char}). Since $\int_{A}\mathbb{E}^{\mathbb{Q}}(E_{a}-X_{a})\mu(da)=0$, this also implies that $\left(\operatorname{\Box}_{a\in A}\varrho_{a}\mu(da)\right)(\mathcal{X})=\int_{A}\varrho_{a}(Y_{a})\mu(da)$.
    \par
    By Lemma \ref{lem:has-lebesgue-equi}, the integral infimal convolution is a convex risk measure with the Lebesgue property. Proposition \ref{prop:jst-subdifferential} thus implies the existence of a subgradient $\mathbb{Q}\in\mathscr{M}_{\mathbb{P}}$ of $\operatorname{\Box}_{a\in A}\varrho_{a}\mu(da)$ at $\mathcal{X}$. We claim $(\mathbb{Q},(Y_{a})_{a\in A})$ is an $\mathscr{E}$-competitive equilibrium. Suppose, for the sake of contradiction, that $(\mathbb{Q},(Y_{a})_{a\in A})$ is not an $\mathscr{E}$-competitive equilibrium.
    \par
    Define a correspondence $F:A\longrightarrow 2^{L^{\infty}(\mathbb{P})}$ by
    \begin{equation*}
        F(a)=\left\{\mathcal{Z}\in L^{\infty}(\mathbb{P}):\varrho_{a}(\mathcal{Z})<\varrho_{a}(Y_{a})\right\}.
    \end{equation*}
    A consequence of Lemma \ref{lem:g-meas} is that $F$ is $\mathscr{A}$-measurable and has a measurable graph. Define a correspondence $\mathfrak{F}:A\longrightarrow 2^{L^{\infty}(\mathbb{P})}$ by
    \begin{equation*}
        \mathfrak{F}(a)=F(a)\cap\left\{\mathcal{Z}\in L^{\infty}(\mathbb{P}):\mathbb{E}^{\mathbb{Q}}(\mathcal{Z})\geq\mathbb{E}^{\mathbb{Q}}(E_{a})\right\}.
    \end{equation*}
    Remark that $\mathfrak{F}$ is a slice correspondence: $\mathfrak{F}=\mathscr{H}_{F}(\mathcal{Y},\xi)$ for $\mathcal{Y}=\frac{d\mathbb{Q}}{d\mathbb{P}}$ and $\xi(a)=\mathbb{E}^{\mathbb{Q}}(E_{a})$. Hence $\mathfrak{F}$ inherits the property of having a measurable graph from $F$ (see Lemma \ref{lem:slice-meas}). If we denote $B=\left\{\mathfrak{F}\neq\emptyset\right\}$, Proposition \ref{prop:meas-select-specific} thus implies that $B\in\mathscr{A}$, and there exists an $\mathscr{A}$-measurable allocation $(\widetilde{Y}_{a})_{a\in A}$ such that $\widetilde{Y}_{a}\in\mathfrak{F}(a)$ for all $a\in B$. Also, since $(\mathbb{Q},(Y_{a})_{a\in A})$ fails to be an $\mathscr{E}$-competitive equilibrium, $\mu(B)>0$.
    \par
    For $n\in\mathbb{N}$, define
    \begin{equation*}
        B_{n}=B\cap\left\{a:\Vert{\widetilde{Y}_{a}}\Vert_{L^{\infty}}\leq n\right\}
    \end{equation*}
    and find some $n_{0}\in\mathbb{N}$ so that $\mu(B_{n_{0}})>0$. Define an $\mathscr{A}$-measurable allocation $(Z_{a})_{a\in A}$ by
    \begin{equation*}
        Z=\mathbf{1}_{B_{n_{0}}}\widetilde{Y}+\mathbf{1}_{A\setminus B_{n_{0}}}Y.
    \end{equation*}
    The allocation $(Z_{a})_{a\in A}$ is easily seen to be Gelfand integrable. Writing $\mathcal{Z}=\int_{A}Z_{a}\mu(da)$, we must have that
    \begin{equation}\label{eq:ineq-q-and-conv}
        \left(\operatorname{\Box}_{a\in A}\varrho_{a}\mu(da)\right)(\mathcal{Z})<\left(\operatorname{\Box}_{a\in A}\varrho_{a}\mu(da)\right)(\mathcal{X})\textrm{ and }\mathbb{E}^{\mathbb{Q}}(\mathcal{Z})\geq\mathbb{E}^{\mathbb{Q}}(\mathcal{X}).
    \end{equation}
    The first inequality follows from the fact that $\int_{A}\varrho_{a}(Y_{a})\mu(da)=\left(\operatorname{\Box}_{a\in A}\varrho_{a}\mu(da)\right)(\mathcal{X})$, but $\varrho_{a}(Z_{a})\leq\varrho_{a}(Y_{a})$ for all $a\in A$, with strict inequality occuring on the non-$\mu$-null set $B_{n_{0}}$, The second inequality follows from the fact that $\mathbb{E}^{\mathbb{Q}}(Z_{a})\geq\mathbb{E}^{\mathbb{Q}}(E_{a})$ for all $a\in B_{n_{0}}$, while $\mathbb{E}^{\mathbb{Q}}(Z_{a})=\mathbb{E}^{\mathbb{Q}}(Y_{a})=\mathbb{E}^{\mathbb{Q}}(E_{a})$ for all $a\in A\setminus B_{n_{0}}$.
    \par
    By the subgradient inequality and (\ref{eq:ineq-q-and-conv}), we have that
    \begin{equation*}
        0>\left(\operatorname{\Box}_{a\in A}\varrho_{a}\mu(da)\right)(\mathcal{Z})-\left(\operatorname{\Box}_{a\in A}\varrho_{a}\mu(da)\right)(\mathcal{X})\geq\mathbb{E}^{\mathbb{Q}}(\mathcal{Z}-\mathcal{X})\geq0,
    \end{equation*}
    which is a contradiction. So $(\mathbb{Q},(Y_{a})_{a\in A})$ is an $\mathscr{E}$-competitive equilibrium, as desired.
\end{proof}
\section*{Acknowledgments}
We would like to express our immense gratitude to two anonymous reviewers, whose cogent suggestions significantly improved the quality of the article.
\printbibliography

@article{melnikov-general-theory,
    author = {Vasily Melnikov},
    title = {A General Theory of Risk Sharing},
    journal = {arXiv:2505.19276v2},
    year = {2025}
}

@article{felixfs,
    author = {Felix-Benedikt Liebrich},
    title = {Risk Sharing Under Heterogeneous Beliefs Without Convexity},
    journal = {Finance and Stochastics},
    year = {2024},
    pages = {999-1033},
    volume = {28}
}

@article{aumanncontinuum,
    author = {Robert Aumann},
    title = {Markets With a Continuum of Traders},
    journal = {Econometrica},
    year = {1964},
    volume = {32},
    pages = {39-50}
}

@article{righi-op,
    author = {Marcelo Righi and Marlon Moresco},
    title = {Inf-Convolution and Optimal Risk Sharing With Countable Sets of Risk Measures},
    journal = {Annals of Operations Research},
    year = {2024},
    volume = {336}
}

@book{aliprantis-inf-dim,
    author = {Charalambos Aliprantis and Kim Border},
    title = {Infinite Dimensional Analysis},
    publisher = {Springer},
    year = {2006}
}

@article{consistency,
    author = {Tiantian Mao and Ruodu Wang},
    title = {Risk Aversion in Regulatory Capital Principles},
    journal = {SIAM Journal on Financial Mathematics},
    year = {2020},
    volume = {11},
    pages = {169-200}
}

@incollection{schtouz,
    author = {Elyès Jouini and Walter Schachermayer and Nizar Touzi},
    title = {Law Invariant Risk Measures Have the Fatou Property},
    booktitle = {Advances in {Mathematical} {Economics}},
    publisher = {Springer},
    year = {2006},
    pages = {49–71}
}

@article{richter,
    author = {Hans Richter},
    title = {Verallgemeinerung eines in der Statistik benötigten Satzes der Maßtheorie},
    journal = {Mathematische Annalen},
    year = {1963},
    volume = {150},
    pages = {85-90}
}

@book{stoch-fin,
    author = {Hans Föllmer and Alexander Schied},
    title = {Stochastic Finance},
    publisher = {de Gruyter},
    year = {2002}
}

@article{quantile-share,
    author = {Paul Embrechts and Haiyan Liu and Ruodu Wang},
    title = {Quantile-Based Risk Sharing},
    journal = {Operations Research},
    year = {2018},
    pages = {936-949},
    volume = {66}
}

@article{edgar,
    author = {Gerald Edgar},
    title = {Measurability in a {Banach} Space},
    journal = {Indiana University Mathematics Journal},
    year = {1977},
    volume = {26},
    pages = {663-677}
}

@article{regsgetarbed,
    author = {Ruodu Wang},
    title = {Regulatory Arbitrage of Risk Measures},
    journal = {Quantitative Finance},
    year = {2016},
    volume = {16},
    pages = {337-347}
}

@article{starr,
    author = {Ross Starr},
    title = {Quasi-Equilibria in Markets with Non-Convex Preferences},
    journal = {Econometrica},
    year = {1969},
    volume = {37},
    pages = {25-38}
}

@article{sun-yannel,
    author = {Yeneng Sun and Nicholas Yannelis},
    title = {Saturation and the Integration of Banach Valued Correspondences},
    journal = {Journal of Mathematical Economics},
    year = {2008},
    volume = {44},
    pages = {861-865}
}

@article{saturated-opt,
    author = {Nobusumi Sagara},
    title = {Relaxation and Purification for Nonconvex Variational Problems in Dual Banach Spaces: The Minimization Principle in Saturated Measure Spaces},
    journal = {SIAM Journal on Control and Optimization},
    year = {2017},
    volume = {55},
    pages = {3154-3170}
}

@article{law-inv-non-conv,
    author = {Peng Liu and Ruodu Wang and Linxiao Wei},
    title = {Is the Inf-Convolution of Law-Invariant Preferences Law-Invariant?},
    journal = {Insurance: Mathematics and Economics},
    year = {2020},
    volume = {91},
    pages = {144-154}
}

@article{solvencyiibad,
    author = {Stefan Weber},
    title = {{Solvency II}, or How To Sweep the Downside Risk Under the Carpet},
    journal = {Insurance: Mathematics and Economics},
    year = {2018},
    pages = {191-200},
    volume = {82}
}

@article{law-inv-util-conv-dual,
    author = {Elyès Jouini and Walter Schachermayer and Nizar Touzi},
    title = {Optimal Risk Sharing for Law Invariant Monetary Utility Functions},
    journal = {Mathematical Finance},
    year = {2008},
    volume = {18},
    pages = {269-292}
}

@article{dualnonmonotone,
    author = {Beatrice Acciaio},
    title = {Optimal Risk Sharing with Non-Monotone
Monetary Functions},
    journal = {Finance and Stochastics},
    year = {2007},
    volume = {11},
    pages = {267-289}
}

@article{heath-ku,
    author = {David Heath and Hyejin Ku},
    title = {Pareto Equilibria With Coherent Measures of Risk},
    journal = {Mathematical Finance},
    year = {2004},
    volume = {14},
    pages = {163-172}
}

@article{solvencyiimf,
    author = {Damir Filipović and Michael Kupper},
    title = {Optimal Capital and Risk Transfers for Group Diversification},
    journal = {Mathematical Finance},
    year = {2007},
    volume = {18},
    pages = {55-76}
}

@article{caprequirfs,
    author = {Felix-Benedikt Liebrich and Gregor Svindland},
    title = {Risk Sharing for Capital Requirements With Multidimensional Security Markets},
    journal = {Finance and Stochastics},
    year = {2019},
    pages = {925-973},
    volume = {23}
}

@article{conv-rep-aumm,
    author = {Fabián Flores-Bazán and Abderrahim Hantoute},
    title = {Convex Representatives of the Value Function and Aumann Integrals in Normed Spaces},
    journal = {SIAM Journal on Optimization},
    year = {2022},
    volume = {32},
    pages = {2773-2796}
}

@article{ghossoubzhu2025,
  author="{Mario} {Ghossoub} and {Michael} {Zhu}",
  journal = "Mathematical Finance",
  title={Efficiency in pure-exchange economies with risk-averse monetary utilities},
  year = "2025",
  note = "Forthcoming"
}

@article{kluv,
    author = {Igor Kluvánek},
    title = {The range of a vector-valued measure},
    journal = {Mathematical Systems Theory},
    year = {1973},
    volume = {7},
    pages = {44-54}
}

@article{khan85,
    author = {Ali Khan},
    title = {On the integration of set-valued mappings in a nonreflexive Banach space, II},
    journal = {Simon Stevin},
    year = {1985},
    volume = {59},
    pages = {257-267}
}

@article{pod-sat,
    author = {Konrad Podczeck},
    title = {On the convexity and compactness of the integral of a Banach space valued correspondence},
    journal = {Journal of Mathematical Economics},
    year = {2008},
    pages = {836-852},
    volume = {44}
}

@article{sch-ivar,
    author = {Ivar Ekeland and Walter Schachermayer},
    title = {Optimal Transport and the Geometry of $L^{1}(\mathbb{R}^{d})$},
    journal = {Proceedings of the American Mathematical Society},
    year = {2014},
    pages = {3585-3596},
    volume = {142}
}

@article{welfare-theorem-second-anderson,
    author = {Robert M. Anderson},
    title = {The Second Welfare Theorem with Nonconvex Preferences},
    journal = {Econometrica},
    year = {1988},
    volume = {56},
    pages = {361-382}
}

@article{leese,
    author = {Stephen Leese},
    title = {Multifunctions of Souslin type},
    journal = {Bulletin of the Australian Mathematical Society},
    year = {1974},
    volume = {11},
    issue = {3},
    pages = {395-411}
}

@article{felix-charge,
    author = {Massimiliano Amarante and Felix-Benedikt Liebrich and Cosimo Munari},
    title = {Uniqueness of Convex-Ranged Probabilities and Applications to Risk Measures and Games},
    journal = {Mathematics of Operations Research},
    year = {2025},
    volume = {50},
    pages = {743-763}
}

@article{tk,
    author = {Amos Tversky and Daniel Kahneman},
    title = {Advances in prospect theory: Cumulative representation of uncertainty},
    journal = {Journal of Risk and Uncertainty},
    year = {1992},
    volume = {5},
    pages = {297-323}
}

@article{subgrad_law_inv,
    author = {Gregor Svindland},
    title = {Subgradients of law-invariant convex risk measures on ${L^{1}}$},
    journal = {Statistics {\&} Risk Modeling},
    year = {2010},
    volume = {27},
    pages = {169-199}
}

@book{kechris,
    author = {Alexander Kechris},
    title = {Classical Descriptive Set Theory},
    publisher = {Springer},
    year = {1995}
}

@book{core-large-econ-book,
    author = {Werner Hildenbrand},
    title = {Core and Equilibria of a Large Economy},
    publisher = {Princeton University Press},
    year = {1974}
}

@article{vind-edge,
    author = {Karl Vind},
    title = {Edgeworth-Allocations in an Exchange Economy with Many Traders},
    journal = {International Economic Review},
    year = {1964},
    volume = {5},
    pages = {165-177}
}

@article{hidden-assumption,
    author = {Rabee Tourky and Nicholas Yannelis},
    title = {Markets with Many More Agents than Commodities: Aumann's “Hidden” Assumption},
    journal = {Journal of Economic Theory},
    year = {2001},
    volume = {101},
    pages = {189-221}
}

@article{lee-comp,
    author = {Sangjik Lee},
    title = {Competitive Equilibrium with an Atomless Measure Space of Agents and Infinite Dimensional Commodity Spaces Without Convex and Complete Preferences},
    journal = {Hitotsubashi Journal of Economics},
    year = {2013},
    volume = {54},
    pages = {221-230}
}

@article{convexity-assumption,
    author = {Michael Farrell},
    title = {The Convexity Assumption in the Theory of Competitive Markets},
    journal = {Journal of Political Economy},
    year = {1959},
    volume = {67},
    pages = {377-391}
}

@article{aumann66,
    author = {Robert Aumann},
    title = {Existence of Competitive Equilibria in Markets with a Continuum of Traders},
    journal = {Econometrica},
    year = {1966},
    pages = {1-17},
    volume = {34}
}

@article{jang-lee,
    author = {Hyo Seok Jang and Sangjik Lee},
    title = {Equilibria in a large production economy with an infinite dimensional commodity space and price dependent preferences},
    journal = {Journal of Mathematical Economies},
    year = {2020},
    pages = {57-64},
    volume = {90}
}

@book{edge-psychic,
    author = {Francis Edgeworth},
    title = {Mathematical Psychics: An Essay on the Application of Mathematics to the Moral Sciences},
    publisher = {Kegan Paul},
    year = {1881}
}

@article{hild-welfare,
    author = {Werner Hildenbrand},
    title = {Pareto Optimality for a Measure Space of Economic Agents},
    journal = {International Economic Review},
    year = {1969},
    volume = {10},
    pages = {363-372}
}

@book{gro-tvs,
    author = {Alexander Grothendieck},
    title = {Topological vector spaces},
    publisher = {Gordon and Breach},
    year = {1973}
}
\end{document}